\documentclass[journal]{IEEEtran}

\usepackage{epstopdf}
\usepackage[draft]{hyperref} 
\usepackage{setspace}
\usepackage{amsmath}
\usepackage{amssymb}
\usepackage{stfloats}
\usepackage{amsthm}
\usepackage{amsfonts}
\usepackage{color}
\usepackage{graphicx}
\usepackage{subfigure}  

\usepackage[flushleft]{threeparttable} 

\usepackage{upgreek}
\usepackage{bm}
\usepackage{nicefrac}

\usepackage{multirow}

\usepackage{cite}
\usepackage{mathtools}
\usepackage{stmaryrd}
\usepackage[mathscr]{euscript}
\usepackage{array}
\usepackage{mathrsfs}

\usepackage{soul}

\newtheorem{theorem}{Theorem}
\newtheorem{lemma}{Lemma}

\begin{document}
\author{  Satish~Mulleti, {\it Member, IEEE} Yonina C. Eldar, {\it Fellow, IEEE}
	\thanks{\scriptsize S. Mulleti is with the Department of Lectrical Engineering, Indian Institute of Technology (IIT) Bombay, India and Y. C. Eldar is with the Faculty of Math and Computer Science, Weizmann Institute of Science, Israel. Email: mulleti.satish@gmail.com, yonina.eldar@weizmann.ac.il}
	\thanks{\scriptsize This project has received funding from Israeli Council for Higher Education (CHE) via the Weizmann Data Science Research Center; European Union’s Horizon 2020 research and innovation program under grant No. 646804-ERC-COG-BNYQ; and the Israel Science Foundation under grant no. 0100101. }
	
}
\title{Modulo Sampling of FRI Signals}
\markboth{Submitted to the IEEE Transactions on Signal Processing}%
{Shell \MakeLowercase{\textit{et al.}}: Bare Demo of IEEEtran.cls for Journals}
\maketitle

\begin{abstract}
	The dynamic range of an analog-to-digital converter (ADC) is critical during sampling of analog signals. A modulo operation prior to sampling can be used to enhance the effective dynamic range of the ADC. Further, sampling rate of ADC too plays a crucial role and it is desirable to reduce it. Finite-rate-of-innovation (FRI) signal model, which is ubiquitous in many applications, can be used to reduce the sampling rate. In the context of modulo folding for FRI sampling, existing works operate at a very high sampling rate compared to the rate of innovation (RoI) and require a large number of samples compared to the degrees of freedom (DoF) of the FRI signal. Moreover, these approaches use infinite length filters that are practically infeasible. We consider the FRI sampling problem with a compactly supported kernel under the modulo framework. We derive theoretical guarantees and show that FRI signals could be uniquely identified by sampling above the RoI. The number of samples for identifiability is equal to the DoF. We propose a practical algorithm to estimate the FRI parameters from the modulo samples. We show that the proposed approach has the lowest error in estimating the FRI parameters while operating with the lowest number of samples and sampling rates compared to existing techniques. The results are helpful in designing cost-effective, high-dynamic-range ADCs for FRI signals.   
\end{abstract}

\IEEEpeerreviewmaketitle

\section{Introduction}
Finite-rate-of-innovation (FRI) signals have few degrees of freedom which aids in sub-Nyquist sampling \cite{eldar_2015sampling}. A popular FRI signal model is one where the signal consists of a linear combination of delayed copies of a known pulse. Such a model is ubiquitous in several time of flight applications such as radar and ultrasound imaging \cite{bajwa_radar, bar_radar,filho201seismic,rip,carter_sonar,eldar_sos,eldar_beamforming}. A typical sampling and reconstruction framework for such signals consists of a tailor-made sampling kernel followed by a sampler or analog to digital converter (ADC) and a parameter estimation block. The sampling kernels are designed to spread the FRI signal information such that by using a low-rate, finite-number of samples, the parameter-estimation block estimates the time-delays and amplitudes \cite{vetterli, eldar_sos, fri_strang, blu_moms, mulleti_kernal}. The key focus of prior works on FRI was to reduce the sampling rate of the ADCs and consequently their cost and power consumption. 

Apart from the sampling rate, another critical parameter of an ADC is its dynamic range. The signal's dynamic range should be well within the ADC's dynamic range. Otherwise, the signal will be clipped, and perfect recovery is not guaranteed from the samples. In the FRI framework, the dynamic range of the signals may vary significantly due to a large variation in the amplitudes of the targets in time-of-flight imaging applications. Hence, the problem of designing an ADC that can sample a wide range of FRI signals without clipping and simultaneously operating at the lowest possible rate is of great importance. This problem is the focus of this paper.

A \emph{modulo} preprocessing step can be used prior to sampling to address the dynamic range issue. The modulo operation folds or wraps the signal to keep it within the ADC's dynamic range prior to sampling. In this way, the ADC can sample signals with a very high dynamic range. Due to the non-linearity of the modulo operation, perfect reconstruction of the original signal from the modulo or wrapped signal samples may not be possible and additional steps may be required. For example, \emph{self-reset} ADCs are discussed in the context of imaging where the signal is wrapped prior to sampling \cite{sradc_park, sradc_sasagwa, sradc_yuan, krishna2019unlimited}. These frameworks use side information, such as the amount of folding for each sample or the folding sign for perfect reconstruction. However, measuring the side information requires complex hardware circuitry.
 
Theoretically and practically, in the absence of side information, one can still recover the original signal from its modulo samples by using unwrapping algorithms. Bhandari et al. \cite{uls_tsp} showed that theoretically, it is possible to reconstruct a bandlimited signal from its modulo samples perfectly, provided that the sampling rate is higher than the Nyquist rate. The authors also proposed an algorithm based on the higher-order differences (HoD) to unwrap the modulo samples up to a constant factor. In this algorithm, the oversampling factor (OF) is $(2\pi e) \approx 17$, where $e$ is the Euler's constant. Romonov and Ordentlich \cite{uls_romonov} proposed an alternative method for unwrapping bandlimited signals by using the fact that few samples of the signal are always within the dynamic range and unaffected by the modulo operation. The folded samples can be estimated from the unfolded ones by using a linear prediction (LP) approach. The algorithm requires that $\text{OF}>1$. Recently, Ayush et al. \cite{bhandari2021unlimited} presented results for periodic bandlimited (PBL) signals. They showed that the folding instants of the modulo signal could be determined by using the out-of-band Fourier coefficients of the modulo signal. The sampling rate is proportional to the number of Fourier coefficients to be determined, which depends on the number of folding instants. Although an upper bound on the number of folding instants is not provided, it increases as the dynamic range of the ADC decreases \cite{bhandari2021unlimited}. Recently, we proposed robust and sampling efficient unfolding algorithm for bandlimited signals \cite{eyar_icassp,eyar_tsp}. The algorithm is shown to operate close to the Nyquist rate compared to existing algorithms for bandlimited signals. 

Although, there are several extensions of modulo sampling for different signal models such as  wavelets \cite{uls_wavelet}, mixture of sinusoids \cite{uls_sinmix}, multi-dimensional signal \cite{uls_md}, finite-dimensional sparse vector \cite{uls_gamp, uls_sparsevec}, sum of exponentials \cite{uls_doa}, computed tomography signals \cite{uls_radon}, graph signals \cite{uls_graph}, there is only a single result for FRI signals \cite{uls_sparse}. In \cite{uls_sparse}, the authors proposed an algorithm to reconstruct FRI signals under the modulo framework by using a periodic lowpass filter (LPF) as the sampling kernel. With this kernel, the filtered signal is a PBL signal which undergoes a modulo operation and is then sampled. The reconstruction process consists of two stages: (i) unfolding of the modulo samples and, (ii) estimation of the FRI parameters from the unfolded samples. For the first stage, the authors rely on the HoD-based approach \cite{uls_tsp} for which the signal is sampled at $(2\pi e)$-times the rate of innovation (RoI) of the FRI signal. After unwrapping, the annihilating filter (AF) method is used for FRI parameter estimation \cite{prony}. Specifically, the Fourier coefficients of the FRI signals are computed from the unfolded samples. Then FRI parameters are estimated from the Fourier samples as in standard Fourier-domain FRI signal reconstruction \cite{eldar_sos, mulleti_kernal}. Ideally, $2L$ unfolded samples should be sufficient for perfect recovery of the FRI signal where $L$ is the number of pulses. However, to invert the HoD operation, the algorithm requires a higher number of samples that are inversely proportional to the dynamic range of the ADC. Hence, in the two-stage recovery framework, the first stage is governed by the OF for unfolding, and the second stage depends on the number of samples for FRI recovery. In \cite{uls_sparse},  the unfolding requires a 17-times higher sampling rate, and FRI recovery requires much higher than $2L$ measurements. Note that unwrapping is unique up to a constant factor. Due to an unknown constant factor in the unwrapped signal, a standard Fourier-domain recovery of the FRI signal may not be possible. This key issue was not addressed in \cite{uls_sparse}.

The unfolding results discussed for bandlimited signals \cite{uls_tsp, uls_romonov}, and for PBL \cite{bhandari2021unlimited} can be extended to FRI signals by using an LPF and a periodic-LPF, respectively. For example, the LPF output of an FRI signal is a bandlimited signal. Hence, unwrapping algorithms of \cite{uls_tsp} or \cite{uls_romonov} can be used in the unfolding step of the two-stage reconstruction process discussed above. However, due to the infinite support of the LPF and the filtered output, countably infinite unfolded samples are required to determine the Fourier samples for FRI recovery. Similarly, one can use a periodic LPF as in \cite{uls_sparse} or a sum-of-sincs (SoS) kernel \cite{eldar_sos, mulleti_kernal} to convert an FRI signal to an equivalent PBL signal. Then the algorithm in \cite{bhandari2021unlimited} can be used for unfolding. However, theoretical identifiability results for PBL signals are not derived in \cite{bhandari2021unlimited}.

From the above survey, we summarize the shortcomings of existing approaches: 
\begin{itemize}
	\item The sampling kernel is a critical component in an FRI sampling framework. Typically, it is desirable to have compactly supported sampling kernels from a practical implementation aspect. However, existing approaches for modulo recovery are based on infinite support sampling kernels.   
	
	\item Theoretical guarantees for uniquely identifying FRI signals from modulo samples are missing, especially for compactly supported kernels. Since identifiability results are independent of any algorithm, they can act as a benchmark to evaluate the efficiency of any algorithm. 
	
	\item Existing algorithms for modulo sampling of FRI signals operate at a much higher sampling rate than the RoI. High sampling rates lead to expensive and power-consuming ADCs and require higher storage and computational cost.
	
	\item Existing algorithms determine the true samples from their modulo ones up to a constant factor. This unknown factor is neglected in the FRI recovery. However, the unknown factor will not allow FRI recovery methods to be applied directly in practice, and an alternate approach is required.  
\end{itemize}

In this paper, we address these shortcomings. We consider the problem of modulo sampling of FRI signals by using a compactly supported SoS kernel. Our first objective is to derive identifiability results independent of any recovery algorithm and then develop an algorithm that is efficient in terms of sampling rate and the number of measurements. In this context, our main contributions are summarized as follows.
\begin{itemize}
	\item We consider modulo sampling with a compactly supported SoS kernel for FRI signals. It is shown that the output of an SoS kernel is a PBL signal \cite{eldar_sos, mulleti_kernal}. The PBL signal is folded by using a modulo operator and then sampled. We show that an $L$-order PBL signal can be uniquely identifiable up to a constant factor from their modulo samples if the number of samples per interval is greater than or equal to $2L+1$ and are prime. We also show that $2L+1$ samples are necessary.
	
	\item To derive the FRI parameters from PBL samples (samples of the output of the SoS kernel), AF can be used. However, the samples are identifiable up to a constant factor, and AF can not be directly applied. To address this issue, we consider three solutions. The first is oversampling where $4L+1$ samples are considered in the first approach. In the second approach, only $2L+1$ samples are used by assuming that the time delays are on a grid. Third approach requires the amplitudes to be positive. This approach does not need oversampling or an on-grid assumption. 
	\item We propose a sampling-efficient algorithm to estimate the FRI parameters from modulo samples. We use Itoh's approach \cite{itoh} for unfolding the samples of the SoS kernel. For unfolding, we derive constraints on the SoS filter coefficients. We then establish bounds on the minimum sampling rate and the number of measurements for unfolding and FRI parameter estimation. In particular, we show that $2L+1$ modulo samples are sufficient for estimating the FRI parameters.  
	\item We present simulation results where the proposed algorithm is compared with existing approaches \cite{uls_tsp, uls_sparse, bhandari2021unlimited}. We show that our method requires the lowest sampling rate and number of samples compared to \cite{uls_tsp, uls_sparse, bhandari2021unlimited}. 
\end{itemize}

The paper is organized as follows. The next section presents the signal model and an explicit problem formulation. In Section \ref{sec:no_modulo}, we discuss the FRI sampling framework in the absence of modulo operation, which will be helpful in understanding the later sections. In Section \ref{sec:identifiability} we present theoretical guarantees for the modulo-FRI framework. A practical algorithm is presented in Section \ref{sec:algorithm}. The proposed algorithm is compared with existing results in Section \ref{sec:comparision}, followed by conclusions.

We use the following notations and symbols throughout the paper. For any $a \in \mathbb{R}$ and $\lambda \in \mathbb{R}^+$, we define a modulo operation $\mathcal{M}_{\lambda}(\cdot)$ as
\begin{equation}
\mathcal{M}_{\lambda}(a) = (a+\lambda)\,\, \text{mod}\,\, 2\lambda -\lambda.	
\label{eq:mod_def}
\end{equation}
For a continuous-time (CT) signal $f(t)$, its uniform samples in CT domain are denoted by $f(nT_s) = \displaystyle f(t) \sum_{m \in \mathbb{Z}} \delta(t-mT_s), n \in \mathbb{Z}$, where $\delta$ is Dirac impulse. In the discrete-domain, the samples are denoted as $f[n] = f(nT_s)$. For a signal $f(t)$ its modulo version is denoted as $f_\lambda(t) = \mathcal{M}_{\lambda}(f(t))$. For any square matrix $\mathbf{A}$, $\sigma(\mathbf{A})$ and $\mathcal{E}(\mathbf{A})$ are the set of eigenvalues and eigenvectors of $\mathbf{A}$, respectively.

\section{Problem Statement}
\label{sec:problme_formulation}
Consider a real-valued FRI signal 
\begin{align}
f(t) = \sum_{\ell=1}^L a_\ell h(t-t_\ell),
\label{eq:fri1}
\end{align}
with the following assumptions:
\begin{enumerate}
	\item[A1] The pulse $h(t)$ is known, real-valued, and time-limited to an interval $[0, T_h]$.
	\item[A2] The number of pulses $L$ is known.
	\item[A3] For $\ell = 1, 2, \cdots, L$, $t_\ell \in (0, T_d] \subset \mathbb{R}$ for known $T_d$.
\end{enumerate}
The objective is to devise a reconstruction framework to recover $\{a_{\ell}, t_{\ell}\}_{\ell=1}^L$ from low-rate or sub-Nyquist samples. This is achieved by designing a sampling kernel $g(t)$ such that from a finite number of sub-Nyquist samples $y(nT_s)$ of the filtered FRI signal $y(t) = (f*g)(t)$, the FRI parameters $\{a_\ell, t_\ell\}_{\ell=1}^L$ are computed uniquely. We discuss the standard sampling and reconstruction approach in the next section. 

\begin{figure*}[!t]
	\centering
	\includegraphics[width= 5in]{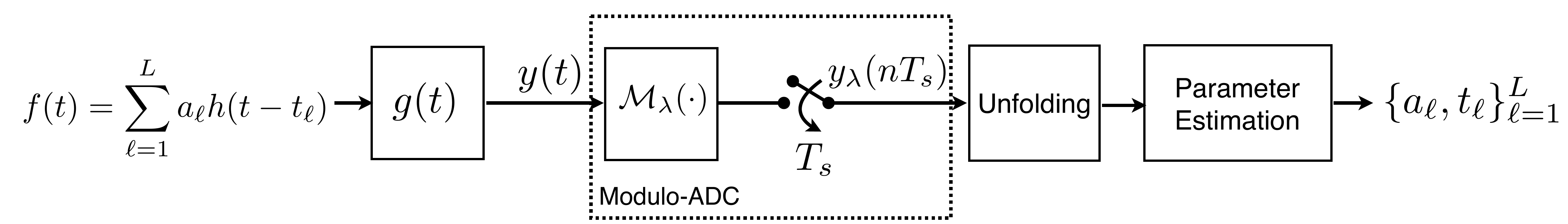}
	\caption{Kernel-based FRI sampling and reconstruction approach: $g(t)$ denotes the sampling kernel; the modulo-ADC consists of a modulo folding operator followed by a conventional ADC with dynamic-range $[-\lambda, \lambda]$; $T_s$ denotes the sampling rate.}
	\label{fig:sampling}
\end{figure*}

In practice, the analog to digital converter (ADC) has a finite dynamic range, $[-\lambda, \lambda]$, and, typically, signals beyond this range are clipped before being sampled. Clipping results in loss of information. One way to overcome clipping is to map the signal to the dynamic range of the ADC. In this work, we consider a modulo operation on the analog signal to restrict to $[-\lambda, \lambda]$ before sampling (cf. Fig.~\ref{fig:sampling}). 

The modulo samples are denoted by $y_{\lambda}(nT_s) = \mathcal{M}_{\lambda}\left(y(nT_s)\right)$.  The objective is to design a sampling kernel and parameter estimation technique such that $\{a_\ell, t_\ell\}_{\ell=1}^L$ are determined from modulo samples $y_{\lambda}(nT_s)$ measured at a sub-Nyquist rate. In addition, it is desirable that the number of samples are close to the number of degrees of freedom (DoF) given by $2L$. In our formulation, we assume that the time-domain signals are real-valued. The framework could be extended to complex-valued signals by modifying the modulo-operator to fold both real and imaginary parts independently. In the next section, we discuss sampling and recovery in the absence of the modulo operation. The discussion will aid in deriving and comparing the results in the presence of modulo.

\section{FRI Sampling and Recovery Without Modulo}
\label{sec:no_modulo}
The FRI sampling and reconstruction problem is posed as a problem of designing a sampling kernel $g(t)$ such that from a finite number of filtered samples of $y(t) = (f*g)(t)$ one can determine the FRI parameters $\{a_\ell, t_\ell\}_{\ell=1}^L$. In particular, the goal is to recover the $2L$ unknowns from samples $\{y(nT_s)\}_{n=1}^N$ such that $N$ is close to $2L$ and the sampling interval,  $T_s$, is as large as possible.  

The design of the sampling kernel is tightly coupled to the reconstruction strategy. Next, we discuss a widely used Fourier-domain recovery method and then examine the sampling kernel that enables this reconstruction \cite{eldar_sos, mulleti_kernal, eldar_2015sampling}.

\subsection{Fourier-Domain Reconstruction}
The Fourier transform of $f(t)$ is given as
\begin{align}
F(\omega) = \int f(t) e^{-\mathrm{j}\omega t} \mathrm{d}t = H(\omega) \sum_{\ell = 1}^L a_\ell \, e^{-\mathrm{j}\omega t_\ell},
\label{eq:F}
\end{align}
where $H(\omega)$ is the Fourier transform of $h(t)$. Next, consider the following samples 
\begin{align}
S(k\omega_0) = \frac{F(k\omega_0)}{H(k\omega_0)} =  \sum_{\ell = 1}^L a_\ell \, e^{-\mathrm{j}k\omega_0 t_\ell}, \quad k \in \mathcal{K},
\label{eq:Skomega}
\end{align}
where $\mathcal{K}$ is a set integers and $\omega_0$ is the sampling interval in the frequency domain. We assume\footnote{The assumption that the spectral samples of the pulse $h(t)$ are non-vanishing is, typically, satisfied by  many practically applied pulses due to their small time-support and large bandwidth.} that $H(k \omega_0) \neq 0$ for $k \in \mathcal{K}$. It can be shown that $2L$ consecutive samples of $S(k\omega_0)$ are sufficient to uniquely identify $\{a_\ell, t_\ell\}_{\ell=1}^L$ if $\omega_0 = \frac{2\pi}{T_d}$ \cite{eldar_2015sampling}.

In practice, AF method can be used to determines $\{a_\ell, t_\ell\}_{\ell=1}^L$ from $\{S(k\omega_0)\}_{k \in \mathcal{K}}$ provided that $|\mathcal{K}| \geq 2L$. As the AF method plays a crucial role in designing an algorithm in the presence of modulo operation, we discuss the AF method at the end of this section.

Given that $2L$ Fourier samples are sufficient to identify the FRI parameters, the next question is how to compute the Fourier samples $\{F(k\omega_0)\}_{k \in \mathcal{K}}$ as $\{H(k\omega_0)\}_{k \in \mathcal{K}}$ are pre-computed from known $h(t)$. In the following, we discuss a sampling kernel-based approach to determine $\{F(k\omega_0)\}_{k \in \mathcal{K}}$ from sub-Nyquist samples of the filtered signal.

\subsection{Sampling Kernel and Sampling}
\label{sec:sampling_kernel}
 Let us consider a sum-of-sincs kernel \cite{eldar_sos} with impulse response
\begin{align}
g(t) = \text{rect}\left(\frac{t}{T_g}\right)\sum_{k\in \mathcal{K}} c_k e^{\mathrm{j}k \omega_0 t},
\label{eq:sos}
\end{align}
where $\text{rect}\left(\frac{t}{T_g}\right)  =1$ for $t \in [0, T_g]$ or zero otherwise. To keep the filter response and its output real-valued we choose $\mathcal{K}$ as
\begin{align}
\mathcal{K} = \{-K, \cdots, K\}, \quad \text{where} \quad K\geq L,
\label{eq:Kset}
\end{align}
and $c_k = c_{-k}^*$. The choice of the set $\mathcal{K}$ implies that $|\mathcal{K}|\geq 2L+1$, that is, one sample more than the DoF $2L$. For $T_g> T_h+T_d$, it can be shown that (cf. \cite{mulleti2021learning}) for $t \in T_{\text{obs}}= [T_h+T_d, T_g]$ the filtered signal $y(t) = (f*g)(t)$ is given as
\begin{align}
y(t) 
 = \sum_{k\in \mathcal{K}} c_k e^{\mathrm{j}k \omega_0 t} \, F(k\omega_0). \label{eq:conv2} 
\end{align}

The filtered signal in $T_{\text{obs}}$  is a linear combination of $\mathcal{K}$ Fourier samples of the input signal $f(t)$. Upon uniform sampling $y(t)$ within the interval $T_{\text{obs}}$ we have
\begin{align}
y(nT_s) = \sum_{k\in \mathcal{K}} c_k e^{\mathrm{j}k \omega_0 nT_s} \, F(k\omega_0), \quad n \in \mathcal{N},  \label{eq:y_F}
\end{align}
where $T_s$ is sampling interval and the set of sampling indices $\mathcal{N}$ is given by
\begin{align}
\begin{aligned}
\mathcal{N} &= \{N_{\min}, \cdots, N_{\max}\}, \\
\text{where  } N_{\min} &= \left\lceil \frac{T_h+T_d}{T_s} \right\rceil, \,\, N_{\max} = \left\lfloor \frac{T_g}{T_s}  \right\rfloor.
\end{aligned}
\label{eq:sample_index}
\end{align}
The relation in \eqref{eq:y_F} denotes set of $|\mathcal{N}|$ linear equations with $|\mathcal{K}|$ unknowns. To compute $|\mathcal{K}|$ Fourier samples it is necessary to have $|\mathcal{N}| \geq |\mathcal{K}|$ time-domain samples which can be insured by setting  $T_g \geq  T_h+T_d + |\mathcal{N}|T_s$. In addition, to uniquely determine $\{F(k\omega_0)\}_{k \in \mathcal{K}}$ from the samples $\{y(nT_s)\}_{n \in \mathcal{N}}$, it is necessary to ensure the elements of the set $\{e^{\mathrm{j}k \omega_0 T_s}\}_{k \in \mathcal{K}}$ are distinct. This condition is satisfied if we have $ |\mathcal{K}|\omega_0 T_s \leq 2\pi$, that is, $T_s \leq \frac{T_d}{|\mathcal{K}|}$. To summarize, the FRI parameters can be computed from the samples $\{y(nT_s)\}_{N_{\min}}^{N_{\max}}$ provided that $T_g \geq  T_h+T_d + |\mathcal{N}|T_s$ and 
\begin{align}
T_s \leq \frac{T_d}{|\mathcal{K}|}.
\label{eq:sampling_interval}
\end{align}
Since $|\mathcal{K}|\geq 2L$, the above result implies that the optimal sampling interval is $T_{s, opt}=\frac{T_d}{2L}$ and RoI is $\frac{2L}{T_d}$. The minimum number of time samples required to identify the FRI parameters is $2L$ which is equal to the DoF of the FRI signal. 
The results are summarized in the following theorem.
\begin{theorem}[Sampling and Reconstruction of FRI Signals]
	\label{thm:FRInomod}
	Consider FRI signals in \eqref{eq:fri1} that satisfy assumptions (A1)-(A3). The FRI parameters can be uniquely recovered from the filtered samples $\{y(nT_s) = (f*g)(nT_s)\}_{n \in \mathcal{N}}$ where the impulse response of the filters $g(t)$ is given as in \eqref{eq:sos} and the sampling set $\mathcal{N}$ as in \eqref{eq:sample_index}, provided that $|\mathcal{N}|\geq |\mathcal{K}| \geq 2L$ and $T_s \leq \frac{T_d}{|\mathcal{K}|}$.
\end{theorem}

The FRI signals have finite support and finite-degrees of freedom over the support. Hence, the conditions for perfect recovery are stated both in terms of the sampling rate and the number of samples. This fact plays an important role in the recovery algorithm of the modulo-based sampling scheme.

\subsection{Annihilating Filter (AF) Approach}
\label{sec:af_approach}
Define a discrete sequence as 
\begin{align}
s[k] = S(k\omega_0) = \frac{F(k\omega_0)}{H(k\omega_0)} =  \sum_{\ell = 1}^L a_\ell \, e^{-\mathrm{j}k\omega_0 \tau_\ell}, \quad k \in \mathcal{K}.
\label{eq:s}
\end{align}
Consider a $L+1$-length sequence $\{x[0], x[1], \cdots, x[L]\}$ and let the convolution $(s*x)[k]$ for $L-K \leq k \leq K$ be
\begin{align}
(s*x)[k] =\sum_{\ell=1}^L a_\ell X(e^{\mathrm{j}\omega_0 \tau_\ell}) e^{-\mathrm{j}k\omega_0 \tau_\ell},
\label{eq:af}
\end{align}
where $X(z)$ is the $z$-transform of the sequence $x[k]$. Consider determining the sequence $x[k]$ such that $(s*x)[k] =0, L-K \leq k \leq K$, that is, a filter that annihilates the sequence $s[k]$. From \eqref{eq:af}, we rewrite the annihilation problem as $\mathbf{V}\mathbf{\tilde{x}} = \mathbf{0}$ where $\mathbf{V}$ is a Vandermonde matrix of size $(2K-L+1) \times L$ given by
\begin{align}
\small
 \begin{pmatrix}
	e^{-\mathrm{j} \omega_0 (L-K)\tau_1} & e^{-\mathrm{j}\omega_0 (L-K)\tau_2} & \hdots & e^{-\mathrm{j} \omega_0 (L-K) \tau_L} \\
	e^{-\mathrm{j} \omega_0 (L-K+1)\tau_1} & e^{-\mathrm{j} \omega_0 (L-K+1)\tau_2} & \hdots & e^{-\mathrm{j}\omega_0 (L-K+1) \tau_L}\\
	\vdots & \vdots & \ddots & \vdots \\
	e^{-\mathrm{j} \omega_0 K\tau_1} & e^{-\mathrm{j} \omega_0 K\tau_2} & \hdots & e^{-\mathrm{j} \omega_0 K \tau_L}
	\end{pmatrix}
\label{eq:af2}
\end{align}
and $\mathbf{\tilde{x}} = [ a_1 X(e^{\mathrm{j}\omega_0 \tau_1}) \,\, a_2 X(e^{\mathrm{j}\omega_0 \tau_2}) \cdots a_L X(e^{\mathrm{j}\omega_0 \tau_L})]^{\mathrm{T}}$.
Since $\tau_\ell$s are distinct, for $K \geq L$, the matrix $\mathbf{V}$ has full column rank. Hence \eqref{eq:af2} holds if and only if $X(e^{\mathrm{j}\omega_0 \tau_\ell}) = 0$ for $\ell=1, 2, \cdots, L$, that is, if the zeros of Fourier transform of $x[k]$ are located at $\{e^{\mathrm{j}\omega_0 \tau_\ell}\}_{\ell=1}^L$. Since $x[k]$ is a sequence of length $L+1$, its has $L$ zeros, and there exist a unique sequence $x[k]$ that annihilates $s[k]$. Once the annihilating filter is determined, $\{t_\ell\}_{\ell=1}^L$ are estimated from its zeros. The annihilating filter $x[k]$ is determined from the sequence $s[k]$ by rewriting $(s*x)[k], L-K \leq k \leq K$ as a set of homogeneous equations $\mathbf{S}_T \mathbf{x} = \mathbf{0}$ where 
\begin{align}
\mathbf{S}_T=\small\begin{pmatrix}
s[L-K] & s[L-K-1] & \hdots & s[-K] \\
\vdots & \vdots & \ddots & \vdots \\
s[0] & s[-1] & \hdots & s[-L] \\
\vdots & \vdots & \ddots & \vdots \\
s[K] & s[K-1] & \hdots & s[K-L]
\end{pmatrix}
\label{eq:af1}
\end{align}
and $\mathbf{x} = [x[0] \,\, x[1]\,\, \cdots x[L]]^{\mathrm{T}}$.
The Toeplitz matrix $\mathbf{S}_T \in \mathbb{C}^{(2K-L+1)\times (L+1)}$ is rank deficient, specifically, for $K \geq L$, it can be shown that $\text{Rank}(\mathbf{S}_T) = L$  for a sequence $s[k]$ that consists of linear combinations of $L$ complex exponentials as shown in \eqref{eq:s} \cite[Proposition 15.2]{eldar_2015sampling}. Hence, $\mathbf{S}_T$ has a unique non-zero null space vector $\mathbf{x}$. 
Once $\{\tau_\ell\}_{\ell=1}^L$ are determined, $\{a_\ell\}_{\ell=1}^L$ are found by solving set of linear equations \eqref{eq:s}.

\section{Identifiability Results With Modulo}
\label{sec:identifiability}
we now present identifiability for modulo sampling of FRI signals. In the presence of a modulo operation (cf. Fig.~\ref{fig:sampling}), the samples are decomposed as
\begin{align}
	y_{\lambda}(nT_s) = y(nT_s) +z(nT_s),
	\label{eq:mod_decompose}
\end{align}
where the values of the sequence $z(nT_s)$ are integer multiples of $2\lambda$. The sequence $z(nT_s)$ is a function of $y(t)$, the input to the modulo block, and insure that $|y_{\lambda}(nT_s)|\leq \lambda$. The samples $y_\lambda(nT_s)$ are functions of the FRI parameters. The question we would like to answer is whether the FRI parameters can be uniquely identifiable from $y_\lambda(nT_s)$.

As earlier, we use the SoS kernel (cf. \eqref{eq:sos}). Hence, $y(t)$ is a trigonometric polynomial as in \eqref{eq:conv2}. From Theorem \ref{thm:FRInomod} we note that $|\mathcal{N}|\geq |\mathcal{K}|\geq 2L$ samples of $y(t)$ are sufficient to uniquely identify the FRI signal provided that $T_s \leq \frac{T_d}{|\mathcal{K}|}$. Hence, we first present the identifiability results of the samples of a trigonometric polynomial from its modulo samples and then extend the results to the FRI case.

\subsection{Uniqueness of Trigonometric Polynomial Under Modulo Operation  }
\label{sec:tp_identifiability}
We first present results for a generic trigonometric polynomial and then show how it can be related to the FRI signal model. Consider a $K$-th order real-valued trigonometric polynomial as in \eqref{eq:conv2}. Let the polynomial be sampled at a rate $T_s = \frac{T_d}{2K^{\prime}+1}$ where $K^{\prime} \geq K$. For simplicity, we assume that $c_k =1$ for $k \in \mathcal{K}$. The samples are given as
\begin{align}
	y(nT_s) = \sum_{k=-K}^K F(k\omega_0) e^{\mathrm{j}k\omega_0 nT_s},
	\label{eq:ynTs}
\end{align}
where $\omega_0 = \frac{2\pi}{T_d}$. Consider the problem of uniquely identifying $y(nT_s)$ from its modulo samples $y_\lambda(nT_s)$.
In this case, our identifiability results are presented in the following theorem.
\begin{theorem}[Identifiability of Trigonometric Polynomial From Modulo Samples] 
	\label{thm:tp_identity}
	Consider the modulo samples $y_\lambda(nT_s)$ of the trigonometric polynomial \eqref{eq:ynTs}, where $T_s = \frac{T_d}{2K^{\prime}+1}$ with $K^{\prime} \geq K$.
	\begin{enumerate}
		\item If $K^{\prime}\leq K$ then $y(nT_s)$ is not identifiable from $y_\lambda(nT_s)$.
		\item  If $K^{\prime} > K$ and $2K^{\prime}+1$ is prime then $y(nT_s)$ is uniquely identifiable up to a constant multiple of $2\lambda$ from its modulo samples $y_{\lambda}(nT_s)$.
	\end{enumerate}
\end{theorem}
 \begin{proof}
 	 We first consider the necessary condition and then prove sufficiency. \\
 	 \emph{Necessary Conditions:} We show that for $K^{\prime}\leq K$ there exist another trigonometric polynomial of order $K$ whose modulo samples coincide with $y_{\lambda}(nT_s)$. We first consider the case when $K^{\prime}< K$. Note that to uniquely determine $\{F(k\omega_0)\}_{k =-K}^K$ from the samples $y(nT_s)$ in \eqref{eq:ynTs} there should be more than or equal to $2K+1$ samples within a time interval of length $T_d$. To this end the sampling interval should satisfy the inequality $T_s \leq \frac{2\pi}{T_d}$. Otherwise, there exists an alternative real-valued trigonometric polynomial 
 	 \begin{align}
 	 	\hat{y}(t) = \sum_{k=- K}^K \hat{F}(k\omega_0)e^{\mathrm{j}k\omega_0 t}
 	 	\label{eq:alty}
 	 \end{align}
 	 such that $y(nT_s) = \hat{y}(nT_s)$. Hence, we have $y_{\lambda}(nT_s) = \hat{y}_{\lambda}(nT_s)$. This proves that for $K^{\prime}< K$ one can not uniquely recovery $y(nT_s)$ from $y_{\lambda}(nT_s)$.
 	 
 	 For $K^{\prime}= K$, although one can recover $\{F(k\omega_0)\}_{k =-K}^K$ from $y(nT_s)$, but we cannot uniquely recover $y(nT_s)$ from $y_{\lambda}(nT_s)$. To show this, we construct an alternative polynomial $\hat{y}(t)$ as 
 	 \begin{align}
 	 \hat{y}(t) = y(t) + 2p\lambda = \sum_{k=- K}^K \hat{F}(k\omega_0)e^{\mathrm{j}k\omega_0 t},
 	 \label{eq:alty2}
 	 \end{align} 
 	 where $p \in \mathbb{Z}$. The coefficients of the alternative polynomial are related to those of $y(t)$ as
 	 \begin{align}
 	 \hat{F}(k\omega_0) = F(k\omega_0) + 2p\lambda \, \delta_K(k),
 	 \end{align}
 	 where $\delta_K$ is the Kronecker impulse. The samples of the alternative polynomial are given as $\hat{y}(nT_s) = y(nT_s) = 2p\lambda$ and hence we have that  $\hat{y}_\lambda(nT_s) = y_\lambda(nT_s)$. \\

     
     \noindent\emph{Sufficient Condition:}  For $K^{\prime}>K$ let there exist an alternate solution as in \eqref{eq:alty} 
     Let us assume that there exists an alternate solution as in \eqref{eq:alty} such that $y_{\lambda}(nT_s) = \hat{y}_{\lambda}(nT_s)$. This implies that there exist samples $d(nT_s) \in  \mathbb{Z}$ such that $y(nT_s)-\hat{y}(nT_s) = 2\lambda d(nT_s)$ or
     \begin{align}
     \sum_{k=-K}^K (F(k\omega_0)-\hat{F}(k\omega_0))e^{\mathrm{j}k\omega_0 nT_s}&= 2\lambda d(nT_s).
     \label{eq:nece}
     \end{align}
    Since $T_s = \frac{T_d}{2K^{\prime}+1}$ there are $2K^{\prime}+1$ samples within an interval $T_d$. Without loss of generality let $n = 0, 1, \cdots, 2K^{\prime}$ in \eqref{eq:nece}. Then by using the inverse discrete Fourier transform we have
     \begin{align}
     	\sum_{n=0}^{2K^{\prime}}& d(nT_s) e^{-\mathrm{j}k\omega_0nT_s} =  \nonumber\\ &\begin{cases}
     		\frac{2K^{\prime}+1}{2\lambda}(F(k\omega_0)-\hat{F}(k\omega_0)), \quad\text{for} -K\leq k \leq K, \\
     		0,\quad  \text{for} -K^{\prime}\leq k < -K \,\, \text{and}\,\, K <k\leq K^{\prime}.
     	\end{cases}
     \end{align} 
    This implies that the polynomial $D(z) =\sum_{n=0}^{2K^{\prime}} d(nT_s) z^{n} $ has roots on the unit circle at $\{e^{-\mathrm{j}k\omega_0T_s} \}_{-K^{\prime}\leq k < -K}$ and  $\{e^{-\mathrm{j}k\omega_0T_s} \}_{K <k\leq K^{\prime}}$. Note that the coefficients of the polynomials are integer valued. Next, we use properties of integer valued polynomials and show that $\{e^{-\mathrm{j}k\omega_0T_s} \}_{k \in \{-K,\cdots, K\} \backslash \{0\}}$ too are zeros of $D(z)$. This implies $F(k\omega_0) = \hat{F}(k\omega_0), k \in \{-K,\cdots, K\} \backslash \{0\}$ and uniqueness is established.
    
    To this end, we first analyze the characteristics of the zeros of $D(z)$. In particular, consider the root $z_{K^{\prime}} = e^{-\mathrm{j} K^{\prime}\omega_0  T_s}$. By substituting $\omega_0 = \frac{2\pi}{T_d}$ and $T_s = \frac{T_s}{2K^{\prime}+1}$,  we have that $z_{K^{\prime}} = e^{-\mathrm{j} K\omega_0  T_s} = e^{-\mathrm{j} \frac{K}{2K^{\prime}+1}}$. This implies that $z_{K^{\prime}}$is a $(2K^{\prime}+1 )$-th root of unity. Moreover, since $2K^{\prime}+1$ is prime, $z_{K^{\prime}}$ is $(2K^{\prime}+1 )$-th primitive root of unity. For any primitive root of unity, there exist a Cyclotomic polynomial $Q(z)$ such that $Q(z_{K^{\prime}}) = 0$ where a Cyclotomic polynomial is a monic polynomial with integer coefficients. Importantly, it is the minimal polynomial over the field of rational numbers of any primitive $n$th-root of unity. Hence, the degree of $Q(z)$ is less than or equal to $D(z)$. From the polynomial remainder theorem we have that 
    \begin{align}
    	D(z) = A(z)Q(z) +R(z),
    	\label{eq:rem_thm}
    \end{align} 
  where $A(z)$ and $R(z)$ are polynomials with integer coefficients and the degree of $R$ is less than that of $Q$. Since $D(z_{K^{\prime}}) = Q(z_{K^{\prime}}) = 0$ from \eqref{eq:rem_thm} we have that $R(z_{K^{\prime}}) = 0$. However, since $Q(z)$ is the minimal polynomial with integer coefficient with root at $z_{K^{\prime}}$, this implies that $R(z) = 0$. Since all the primitive roots of unity are zeros of the corresponding Cyclotomic polynomial, $Q(z)$ has zeros $\{e^{-\mathrm{j}\frac{k}{2K^{\prime}+1}}\}_{k=-K^{\prime}}^{-1}$. This implies that $F(k\omega_0) = \hat{F}(k\omega_0)$ for $k = -K, \cdots,-1, 1,\cdots,  K$. In addition, we have that $F(0) \neq  \hat{F}(0)$. In particular from \eqref{eq:nece} we conclude that $F(0) - \hat{F}(0) \in 2\lambda \mathbb{Z}$.  This implies that the trigonometric polynomials $y(nT_s)$ and $\hat{y}(nT_s)$ are identical up to a constant factor which is a multiple of $2\lambda$. 
  \end{proof}
 
 \begin{figure}[!t]
 	\centering
 	\includegraphics[width= 3in]{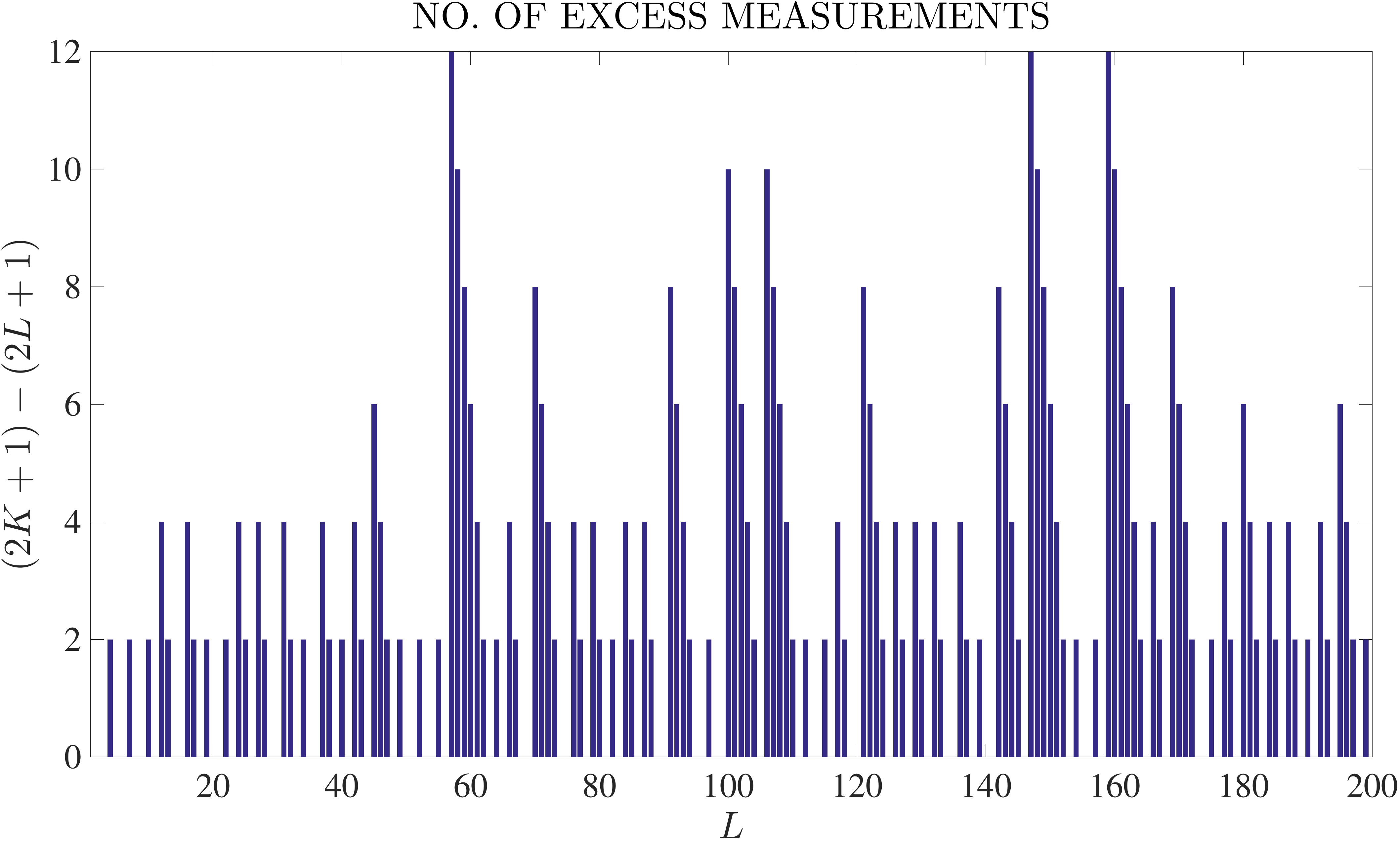}
 	\caption{Number of excessive samples required above the optimum $2L+1$ with the prime condition:}
 	\label{fig:primeof}
 \end{figure}
 In the FRI framework, Theorem~\ref{thm:FRInomod} shows that from the filtered samples $y(nT_s)$ (as in \eqref{eq:y_F}), the FRI signal can be uniquely recovered without modulo provided that $T_s \leq \frac{T_d}{|\mathcal{K}|}$ where $|\mathcal{K}|\geq 2L+1$. With modulo operation prior to sampling, we conclude that $y(nT_s)$ in \eqref{eq:y_F} can be uniquely identifiable from its modulo samples up to a constant factor provided that $T_s \leq \frac{T_d}{|\mathcal{K}|}$,  $K \geq L$, and the number of Fourier samples is prime, that is, $2K+1$ is prime. The prime condition on the number of Fourier samples, which is also equal to the number of time samples of $y(t)$ within an interval of length $T_d$, results in oversampling. For example, for $L=4$, the optimum number of Fourier samples in the absence of modulo operation is $9$, whereas, with modulo operation it is $11$, two samples more than minimum. In Fig.~\ref{fig:primeof}, we plot the difference between the desired number of Fourier samples for unique identifiability and the optimum number of samples $2L+1$ for $2 \geq L \geq 200$. We observe that for $L \leq 56$ one requires to measure maximum of 6 more samples than the optimal. For $57 \leq L \leq 200$ a maximum of 12 addition samples are required for certain values of $L$. In addition, for several values of $L$, no additional samples are required. With this observation, we infer that the prime condition does not warrant high oversampling. 

 \begin{figure*}[!t]
 	\begin{align}
 		\underbrace{\begin{pmatrix}
 				\bar{s}[0] & \bar{s}[-1] & \hdots & \bar{s}[-L] \\
 				\bar{s}[1] & \bar{s}[0] & \hdots & \bar{s}[-L+1] \\
 				\vdots & \vdots & \ddots & \vdots \\
 				\bar{s}[L] & \bar{s}[L-1] & \hdots & \bar{s}[0] 
 		\end{pmatrix}}_{\mathbf{\bar{S}}} = 
 		\underbrace{\begin{pmatrix}
 				s[0] & s[-1] & \hdots & s[-L] \\
 				s[1] & s[0] & \hdots & s[-L+1] \\
 				\vdots & \vdots & \ddots & \vdots \\
 				s[L] & s[L-1] & \hdots & s[0] 
 		\end{pmatrix}}_{\mathbf{S}} + \bar{\beta} \underbrace{\begin{pmatrix}
 				1 & 0 & \hdots & 0 \\
 				0 & 1 & \hdots & 0 \\
 				\vdots & \vdots & \ddots & \vdots \\
 				0 & 0 & \hdots & 1
 		\end{pmatrix}}_{\mathbf{I}_{L+1}}
 		\label{eq:toeplitz}
 	\end{align}
 \end{figure*}
\subsection{Uniqueness of FRI Signal Under Modulo Operation }
\label{sec:new_af}
In the previous section, we showed that trigonometric polynomials of the form \eqref{eq:y_F} or \eqref{eq:ynTs} can be uniquely recovered from its modulo samples $y_\lambda(nT_s)$ up to a constant factor. In this section, we derive identifiability results for FRI signals from the recovered samples 
\begin{align}
	\bar{y}(nT_s) = y(nT_S)+\beta, \quad n= N_{\min}, \cdots, N_{\max}-1,
	\label{eq:bary}
\end{align}
where $\beta \in 2\lambda \mathbb{Z}$ is an unknown constant parameter.

By substituting $y(nT_s)$ from \eqref{eq:y_F} into \eqref{eq:bary} we have
\begin{align}
	\bar{y}(nT_s) = \sum_{k\in \mathcal{K}} c_k e^{\mathrm{j}k \omega_0 nT_s} \, F(k\omega_0)+\beta.
	\label{eq:bary2}
\end{align}
From the known $\{c_k\}_{k \in \mathcal{K}}$ and the above set of linear equations we compute the Fourier samples $\{\bar{F}(k \omega_0)\}_{k \in \mathcal{K}}$ where
\begin{align}
	\bar{F}(k \omega_0) = \begin{cases}
		F(k \omega_0), \quad k \in \mathcal{K} \backslash \{0\}, \\
		F(0)+\beta, \quad k = 0.
	\end{cases}
	\label{eq:barf}
\end{align}
Here we assumed that there are sufficient number of samples, that is, $N_{\max}-N_{\min} \geq |\mathcal{K}|$. As in \eqref{eq:s}, we construct a sequence $\bar{s}[k]$ defined as 
\begin{align}
	\bar{s}[k]  \triangleq \frac{\bar{F}(k\omega_0)}{H(k\omega_0)}&
	= s[k]+\bar{\beta} \, \delta[k] \nonumber \\
	&= \sum_{\ell = 1}^L a_\ell \, e^{-\mathrm{j}k\omega_0 \tau_\ell}+\bar{\beta}\, \delta[k], \quad k \in \mathcal{K},
	\label{eq:bars}
\end{align}
where $\delta[k]$ is the Kronecker impulse and $\bar{\beta} = \frac{\beta}{H(0)} $. When $\beta$ is zero, one can uniquely identify the FRI parameters $\{a_\ell, \tau_\ell\}_{\ell=1}^L$ from $2L$ consecutive samples of $\bar{s}[k]$ (see Section \ref{sec:af_approach}). However, for a nonzero $\beta$ uniqueness is not clear.

To have a unique solution, a possible approach is to neglect the sample at $k=0$ and determine the FRI parameters from the remaining samples of $\bar{s}[k]$. With the missing zeroth sample, we first show that, in general, uniqueness is not guaranteed when one considers an optimum number of Fourier measurements ($K=L$). Then we discuss a few approaches to identify the FRI parameters with additional constraints.

\subsubsection{Non-identifiability with missing sample for $K=L$}
Here we show that for $K=L$ one can not uniquely identify $\{a_\ell, \tau_\ell\}_{\ell=1}^L$ from $\{\bar{s}[k]\}_{k = \mathcal{K}\backslash \{0\}}$ (cf. \eqref{eq:bars}) for $\beta \neq 0$. Specifically, there exist an alternative set of FRI parameters $\{\hat{a}_\ell, \hat{\tau}_\ell\}_{\ell=1}^L$ such that
\begin{align}
\sum_{\ell = 1}^L a_\ell \, e^{-\mathrm{j}k\omega_0 \tau_\ell} = \sum_{\ell = 1}^L \hat{a}_\ell \, e^{-\mathrm{j}k\omega_0 \hat{\tau}_\ell}, \quad k = \mathcal{K}\backslash \{0\},
\label{eq:alt1}
\end{align} 
where $K=L$. We show that \eqref{eq:alt1} holds for a specific example. Let 
\begin{align}
\tau_\ell= \frac{T_d}{2L}\ell, \quad \text{and} \quad
\hat{\tau}_\ell= \frac{T_d}{2L}(\ell+L), \quad \ell = 1,\cdots, L,
\label{eq:alt2}
\end{align}
and
\begin{align}
a_\ell= 1, \quad \text{and} \quad
\hat{a}_\ell= -1, \quad \ell = 1,\cdots, L.
\label{eq:alt3}
\end{align}
Then by substituting these time delays and amplitudes in 
\begin{align}
\sum_{\ell = 1}^L a_\ell \, e^{-\mathrm{j}k\omega_0 \tau_\ell} - \sum_{\ell = 1}^L \hat{a}_\ell \, e^{-\mathrm{j}k\omega_0 \hat{\tau}_\ell},
\label{eq:alt4}
\end{align}
we have
\begin{align}
\sum_{\ell = 1}^L a_\ell \, e^{-\mathrm{j}k\omega_0 \tau_\ell} - \sum_{\ell = 1}^L \hat{a}_\ell \, e^{-\mathrm{j}k\omega_0 \hat{\tau}_\ell}= \sum_{\ell = 1}^{2L}  e^{-\mathrm{j}k \frac{2\pi}{2L}}.
\label{eq:alt5}
\end{align}
The above, right-hand-side sum is zero for $-L \leq k \leq L$ except for $k=0$. This implies that \eqref{eq:alt1} holds true for these set of FRI parameter. Hence for $K=L$, identifiability is not achieved with ambiguity at $k=0$. To address this issue, oversampling or constraints on the FRI parameters can be used as discussed in the following.

\subsubsection{Oversampling for unique identifiability with missing sample}
 The FRI parameters can be uniquely determined from  any $2L$ consecutive samples of $\bar{s}[k]$ that does not include the zeroth sample. This condition holds for $K \geq 2L$ and as a consequence annihilating filter can be applied to either the sequence $\{\bar{s}[k]: 1 \leq k \leq K\}$ or to the sequence $\{\bar{s}[k]: -K \leq k \leq -1\}$ to uniquely determine the time delays $\{\tau_\ell\}_{\ell=1}^L$. However, this approach requires twice the number of samples compared to the case when $\beta = 0$. 

Next, we discuss approaches where oversampling is not required and constraints on the FRI parameters are used for identifiability. In this context, we first consider an existing approach (with zeroth sample missing) where the time-delays are assumed to be on a grid and then propose a method where the amplitudes are assumed to be positive and the zeroth sample is taken into consideration.

\subsubsection{On-Grid time delays with missing sample}
In \cite{namman2021fri}, the problem of recovering FRI parameters from Fourier measurements is considered where the zeroth frequency sample not measured. For recovery of the parameters, the authors considered either $K \geq 2L$ measurements with double the sampling rate as discussed in the previous paragraph or assumed that the time delays are on a grid. In the latter case, with on-grid time delays, $K \geq L$ Fourier measurements, with missing zeroth Fourier measurement, are shown to be sufficient for recovering the FRI parameters.

\subsubsection{Positive amplitudes constraint without a missing sample}
While on-grid time delays are a strong assumption in a practical scenario, an alternative approach was presented in the early 1900s. In this approach, Carath\'eodory showed that a positive ($a_\ell  >0, \ell = 1, \cdots, L$) linear combination of $L$ complex exponentials is uniquely determined from its zeroth sample and any other $2L$ samples \cite{caratheodory}. The Carath\'eodory's result requires to have the zeroth sample, whereas we have ambiguity in this sample. In what follows we show that with the assumption $a_\ell >0, \ell = 1, \cdots, L$ one can uniquely recover parameters $\{a_\ell, \tau_\ell\}_{\ell=1}^L$ even with an unknown ambiguity.

For $K = L$,  the Toeplitz matrix $\mathbf{\bar{S}}$ (as in \eqref{eq:af1}) constructed from the sequence $\bar{s}[k]$ is related to true samples as in \eqref{eq:toeplitz}.
It can be shown that $\text{Rank}(\mathbf{S}) = L$ and hence, the time delays can be determined uniquely from its null-space vector $\mathbf{x}$. However, the annihilating-filter $\mathbf{x}$ is not necessarily in the null-space of $\mathbf{\bar{S}}$. Despite that we show that $\mathbf{x}$ can be uniquely determined from $\mathbf{\bar{S}}$ by using the fact that the amplitudes are positive. 
To this end, we rely on following lemma.
\begin{lemma}
	\label{lem:eigvec}
	The matrices $\mathbf{\bar{S}}$ and  $\mathbf{S}$ have same eigenvectors, that is, $\mathcal{E}(\mathbf{\bar{S}}) = \mathcal{E}(\mathbf{S})$, and $\sigma(\mathbf{\bar{S}}) = \bar{\beta} + \sigma(\mathbf{S})$. 
\end{lemma}
\begin{proof}
	Consider any eigenvalue $\gamma \in  \sigma(\mathbf{S})$ with corresponding eigenvector $\mathbf{c} \in \mathcal{E}(\mathbf{S})$ such that $\mathbf{S}\, \mathbf{c} = \gamma \, \mathbf{c}$. By using the equality $\mathbf{S} = \mathbf{\bar{S}} - \bar{\beta} \mathbf{I}_{L+1}$ (cf. \eqref{eq:toeplitz}), we have that $\mathbf{\bar{S}} \mathbf{c} = (\bar{\beta}+\gamma) \mathbf{c}$. This implies that if $\mathbf{c} \in \mathcal{E}(\mathbf{S})$ then $\mathbf{c} \in \mathcal{E}(\mathbf{\bar{S}})$. In addition, if $\gamma \in  \sigma(\mathbf{S})$ then $\gamma + \bar{\beta} \in  \sigma(\mathbf{\bar{S}})$.
	
	By applying a similar approach, it can be verified that if $\mathbf{\bar{c}} \in \mathcal{E}(\mathbf{\bar{S}})$ then $\mathbf{\bar{c}} \in \mathcal{E}(\mathbf{S})$ and hence $\mathcal{E}(\mathbf{\bar{S}}) = \mathcal{E}(\mathbf{S})$. Further, if $\mathbf{\bar{S}} \mathbf{\bar{c}} = \bar{\gamma} \mathbf{\bar{c}}$ for any $\bar{\gamma} \in  \sigma(\mathbf{\bar{S}})$ then  $\mathbf{S}\mathbf{\bar{c}} = (\bar{\gamma} - \bar{\beta})\mathbf{\bar{c}}$ which implies that for every $\bar{\gamma} \in  \sigma(\mathbf{\bar{S}})$ we have $\bar{\gamma} - \bar{\beta} \in \sigma(\mathbf{S})$ and hence $\sigma(\mathbf{\bar{S}}) = \bar{\beta} + \sigma(\mathbf{S})$. 
\end{proof}
Lemma~\ref{lem:eigvec} suggests that the annihilating filter $\mathbf{x}$, which is in $\mathcal{E}(\mathbf{S})$, is also an eigenvector of $\mathbf{\bar{S}}$. Specifically, we have $\mathbf{\bar{S}}\mathbf{x} = \bar{\beta} \mathbf{x}$. The question is how to uniquely identify $\mathbf{x}$ from the eigenvectors of $\mathbf{\bar{S}}$. To this end, we use the condition that the amplitudes of the FRI signal is positive. By using Carath\'eodory-Toeplitz theorem (see \cite[Sec. 15.2.3]{eldar_2015sampling}, \cite[Sec. 4.9.2]{stoica}) it can be shown that $\mathbf{S}$ is positive semidefinite if $a_\ell >0$.  
Hence, we have that $0 \in \sigma(\mathbf{S})$ as $\mathbf{S}\mathbf{x} = \mathbf{0}$  and the remaining eigenvalues are positive. Combining this results with those in Lemma~\ref{lem:eigvec} we conclude that $\bar{\beta}$ is the minimum eigenvalue of $\mathbf{\bar{S}}$. Hence, the eigenvector of $\mathbf{\bar{S}}$ associated with its minimum eigenvalue is the desired annihilating filter $\mathbf{x}$. Since the annihilating filter $\mathbf{x}$ is unique for a given set of FRI parameters, the FRI parameters can be uniquely determined from $\bar{s}[k]$.

\begin{figure}[!t]
	\centering
	\includegraphics[width= 2.1in]{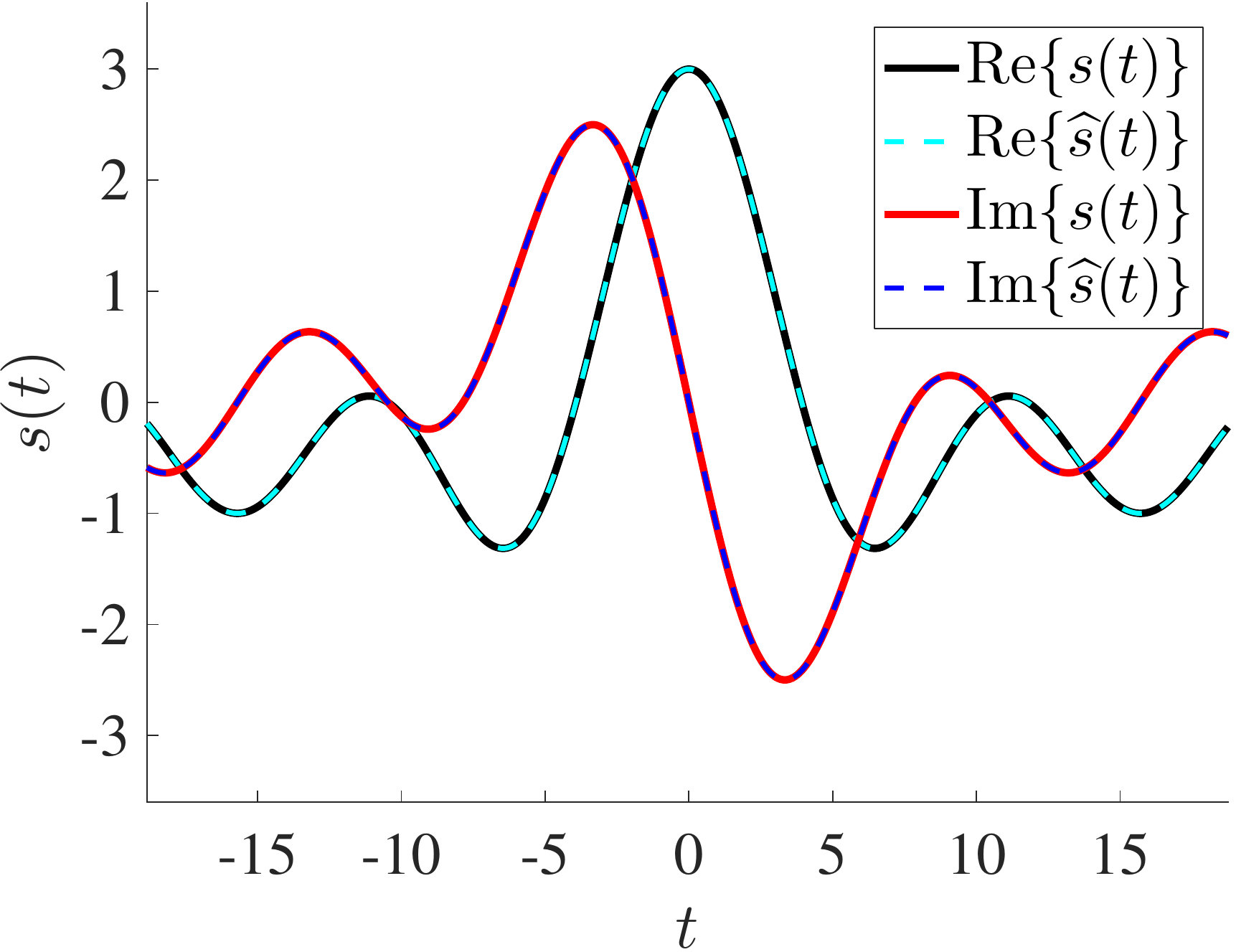}
	\caption{Reconstruction of sum-of-complex-exponential signal $s(t)$ from its uniform samples with a missing sample. We observe perfect reconstruction by using the ASAF algorithm. }
	\label{fig:ASAF}
\end{figure}

We call this approach as the ambiguous sample annihilating filter (ASAF) method.  Note that the ambiguous sample need not be the zeroth one. Lemma~\ref{lem:eigvec} holds for any SoE sequence consisting of $L$ frequencies with positive amplitudes provided that the sample index set $\mathcal{K}$ has at least $2L+1$ consecutive integers and $(L+1)$-th sample is either missing or has measurement uncertainty. To verify our theory we consider $2L+1$ uniform samples of sum-of-complex-exponential signal $s(t) =\displaystyle \sum_{\ell=1}^L a_\ell \, e^{\mathrm{j} \omega_\ell t}$ where $a_\ell >0$. In this particular example, we set $L = 3$, $a_\ell = 1$, and frequencies as  $2\pi[0.2, 0.4, 0.6]$. We consider the problem of recovering the frequencies and amplitudes from $s(nT_s)$, $n \in \{-L, \cdots, -1, 1,\cdots, L\}$ where $T_s =1$. The real and imaginary values of the original signal $s(t)$ and estimated signal $\widehat{s}(t)$ are shown in Fig.~\ref{fig:ASAF}. We observe perfect reconstruction using the ASAF method.

We summarize identifiability results for FRI signals from modulo samples in the following theorem.

\begin{figure}[!h]
	\centering
	\includegraphics[width=2 in]{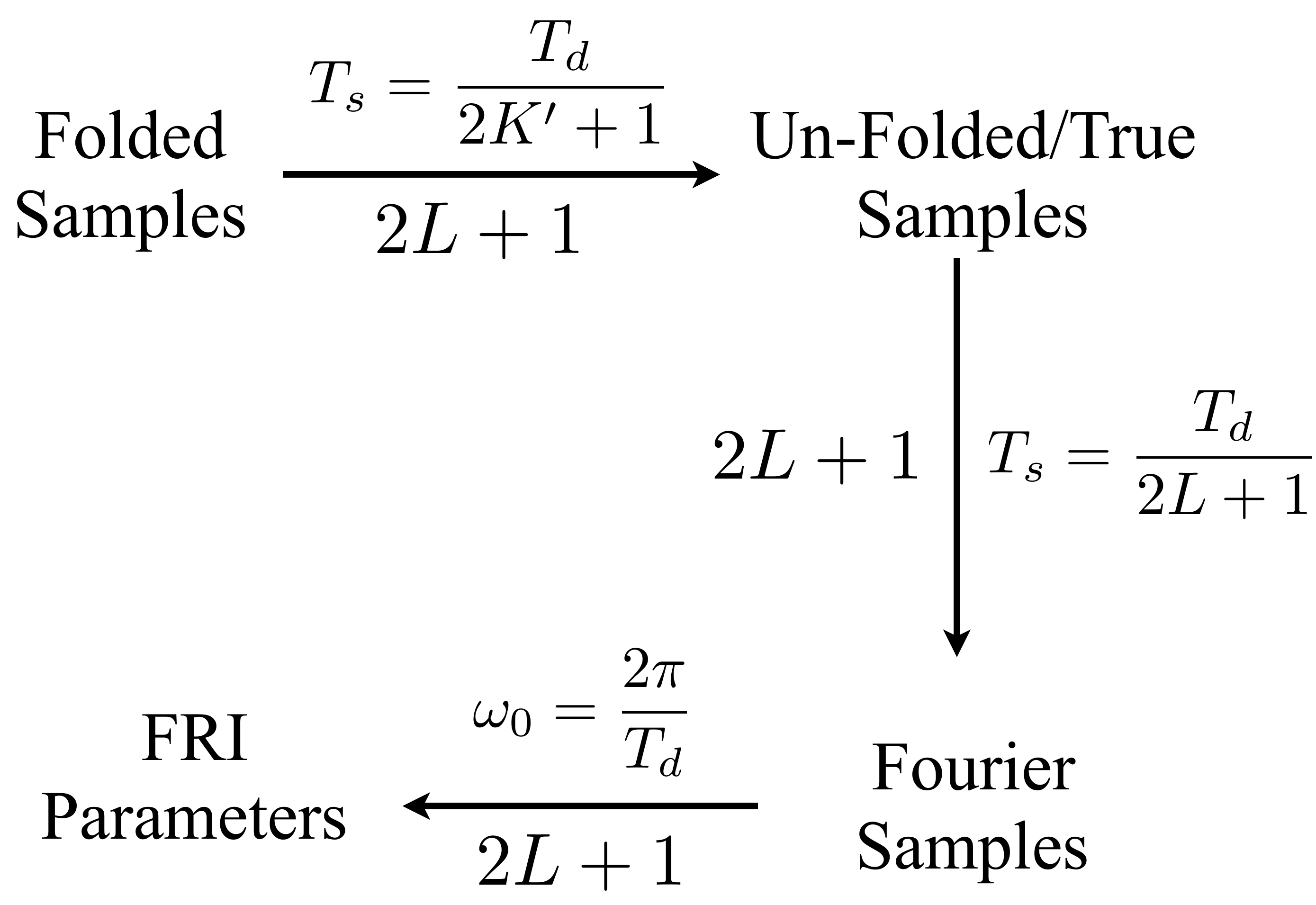}
	\caption{Flow diagram of FRI parameter estimation from folded samples: Sampling intervals and the number of samples for each step are added. For sufficient condition we need $K^{\prime}\geq L$ and $2K^{\prime}+1$ should be prime.}
	\label{fig:Flow}
\end{figure}

\begin{theorem}
	\label{thm:FRImod_iden}
	Consider an FRI signal $f(t)$ as in \eqref{eq:fri1} that follows assumptions (A1) - (A3). Consider a SoS sampling kernel $g(t)$ as in \eqref{eq:sos} whose coefficients satisfy the inequality in \eqref{eq:sos_coeff} where $\mathcal{K} = \{-K, \cdots, K\}$ and $\lambda$ is a known non-zero real number. Consider the modulo samples $\{y_\lambda(nT_s) = \mathcal{M}_\lambda ((f*g)(nT_s))\}_{n \in \mathcal{N}}$, where $T_s \leq \frac{T_d}{2K^{\prime}+1}$ and $K^{\prime} \geq K \geq L$. The sample index set $\mathcal{N}$ is defined in \eqref{eq:sample_index} for $|\mathcal{N}| \geq 2K+1$. Then 
	\begin{enumerate}
		\item FRI signal can not be uniquely identifiable from the modulo samples if $K^{\prime} \leq K$.
		\item FRI signal is uniquely identifiable from the modulo samples if $K^{\prime}>K$ and $2K^{\prime}+1$ is prime and one of the following holds
		\begin{enumerate}
		 \item $K \geq 2L$.
		 \item The time delays are on a grid and $K \geq L$.
		 \item The amplitudes are positive and $K \geq L$.
		\end{enumerate}
	\end{enumerate}
\end{theorem}
To understand the results of Theorem~\ref{thm:FRImod_iden} in comparison to Theorem~\ref{thm:FRInomod} (no modulo folding), we consider the process of determining the FRI parameters from the modulo samples as shown in Fig.~\ref{fig:Flow}. The process has three steps: (1) Unfolding; (2) Determining the Fourier samples from the unfolded or true samples; and (3) Estimating the FRI parameters from the Fourier samples. In Section I, we combined steps 2 and 3 as FRI recovery for ease of explanation, but here we separated them to have a better understanding. In the process shown in Fig.~\ref{fig:Flow}, we assumed that the time-delays are on a grid or the amplitudes are positive. We added the maximum sampling intervals and a minimum number of samples required at each step. Note that the parameters, sampling interval, and the number of samples are crucial for unique recovery as we deal with compactly supported signals with finite DoF. For example, to uniquely determine the Fourier measurements $\{F(k\omega_0)\}_{k=-K}^K$ from the samples $y(nT_s)$ in \eqref{eq:ynTs}, we need a minimum of $2K+1$ samples with maximum sampling interval $T_s = \frac{T_d}{(2K+1)}$. 

Both without folding and with folding approaches require a minimum of $2L+1$ Fourier measurements to determine the FRI parameters uniquely. In both the scenarios, sampling interval in the Fourier-domain is $\omega_0 = \frac{2\pi}{T_d}$. Next, to compute the Fourier measurements from the unfolded or true time samples, a minimum of $2L+1$ time samples are required with sampling interval $T_s = \frac{T_s}{2L+1}$. Up to this point, both approaches are identical in terms of the number of samples and sampling rates. In case of modulo sampling, to unfold $2L+1$ samples, $2L+1$ folded samples are sufficient but the sampling interval is $T_s = \frac{T_s}{2K^{\prime}+1}$ where $K^{\prime}\geq L$ and $2K^{\prime}+1$ is prime. In terms of the number of samples, all the steps require $2L+1$ samples which are optimum. The sampling intervals in steps 2 and 3 in both approaches too are optimum. However, in modulo sampling, unfolding may require oversampling for some values of $L$. Specifically, oversampling is not required for the values of $L$ for which $2L+1$ is prime. In this case $K^{\prime} = L$.

The identifiability results in Theorem~\ref{thm:FRImod_iden} show that without any constraints on the FRI parameters, one requires twice the number of samples. Whereas, with constraints either on the amplitudes or time-delays, identifiability is achieved with optimal number of samples.
Unlike the identifiability results for bandlimited signals \cite{uls_tsp} and periodic bandlimited signals, FRI signals are uniquely identifiable without any constant additive factor. 
Next, we discuss a practical algorithm to recover FRI parameters from the modulo samples.


\section{Filter Design And Recovery Algorithm }
\label{sec:algorithm}
 In this section, we propose an algorithm to estimate the FRI parameters from the modulo samples. We consider an SoS kernel-based modulo sampling framework as discussed in the previous section. Our algorithm follows a two-stage approach where in the first stage unfolding is performed, and in the second stage, FRI parameters are estimated from the unfolded samples. For unfolding, we apply Itoh's first order difference approach \cite{itoh}. The signal reconstruction step is based on the ASAF algorithm. The unfolding step depends on the variation of the filtered signal, and hence it strongly depends on the choice of the SoS filter coefficients. In the following, we first discuss the unfolding and reconstruction and then present an approach to design the SoS filter.

\subsection{Unfolding and Signal Reconstruction}
Our unfolding approach is based on Itoh's results on phase unwrapping \cite{itoh}. The key steps in this method are stated as follows. The modulo samples can be decomposed as
\begin{align}
y_{\lambda}(nT_s) = y(nT_s) +z(nT_s),
\label{eq:mod_decompose1}
\end{align}
where the values of the sequence $z(nT_s) \in 2\lambda \mathbb{Z}$. If the true samples satisfy the inequality 
\begin{align}
|y((n+1)T_s)-y(nT_s)| \leq \lambda
\label{eq:y_lip}
\end{align}
then the following equality hold
\begin{align}
\begin{aligned}
d(nT_s) & \triangleq \mathcal{M}_{\lambda} \left( y_{\lambda}((n+1)T_s)-y_{\lambda}(nT_s) \right),  \\& = y((n+1)T_s)-y(nT_s),
\end{aligned}
\label{eq:diff_samples}
\end{align}
for $n= N_{\min}, \cdots, N_{\max}-1$. This implies that the difference of true samples $y((n+1)T_s)-y(nT_s)$ can be determined from the modulo samples provided that \eqref{eq:y_lip} holds. From the differences, the true samples $y(nT_s)$ are determined up to an unknown constant factor. We can use the annihilating filter approach discussed in Section \ref{sec:new_af} to determine the time delays and amplitudes of the FRI signal. Hence, for unfolding and FRI recovery we need to ensure that \eqref{eq:diff_samples} is satisfied. Note that $y(t)$ is a function of the FRI signal and the sampling kernel. Since one cannot modify the FRI signal, we discuss a approach to design the sampling kernel such that \eqref{eq:diff_samples} holds.

\subsection{Filter Design}
\label{sec:filter_design}
For the filter design problem, we consider a modification to the SoS filter  given in \eqref{eq:sos}. To this end, we first derive a general condition on the filter's impulse response and then discuss the design of the SoS filter coefficients such that the resulting filter satisfies the conditions.

Recall that
\begin{align}
y(t) = \sum_{\ell=1}^L a_\ell \, (h*g)(t-t_\ell)
\end{align}
and $\displaystyle (h*g)(t) = \int h(\tau) g(t-\tau) \mathrm{d}\tau$. Therefore
\begin{align}
& \left | y((n+1)T_s)-y(nT_s) \right| =  \nonumber
\end{align}
\begin{align}
& \left | \sum_{\ell=1}^L a_\ell \left( (h*g)((n+1)T_s-t_\ell)-(h*g)(nT_s-t_\ell) \right)\right|, \nonumber 
\end{align}
\begin{align}
&\leq  \sum_{\ell=1}^L |a_\ell| \left |  (h*g)((n+1)T_s-t_\ell)-(h*g)(nT_s-t_\ell) \right|, \nonumber
\end{align}
\begin{align}
&\leq      \left |  (h*g)((n+1)T_s)-(h*g)(n T_s) \right|       \sum_{\ell=1}^L |a_\ell|, \nonumber
\end{align}
\begin{align}
&=      \left | \int h(\tau) \left(g((n+1)T_s-\tau) -g(nT_s-\tau) \right)\mathrm{d}\tau  \right|       \sum_{\ell=1}^L |a_\ell|, \nonumber
\end{align}
\begin{align}
&\leq       \int |h(\tau)|  \left(|g((n+1)T_s-\tau) -g(nT_s-\tau)|  \right)\mathrm{d}\tau        \sum_{\ell=1}^L |a_\ell|, \nonumber
\end{align}
\begin{align}
&\leq       L a_{\max} \mathcal{L}_{g}(T_s) \int |h(\tau)| \mathrm{d}\tau,   
\label{eq:filter_condition}
\end{align}
where $| g((n+1)T_s-\tau) -g(nT_s-\tau)| \leq \mathcal{L}_{g}(T_s)$. 
Since $L, a_{\max}$ and $\alpha_h=\int |h(\tau)| \mathrm{d}\tau$ are known \emph{a priori}, if the filter $g(t)$ is designed such that for a desired $T_s$ we have
\begin{align}
\mathcal{L}_g(T_s) \leq \frac{\lambda}{L a_{\max} \alpha_h},
\end{align}
then \eqref{eq:y_lip} holds and hence we can estimate $y(nT_s)$ from the modulo samples up to a constant factor. 

Next, consider a SoS filter with impulse response as in \eqref{eq:sos}. The filter is parameterized by the coefficients $\{c_k\}_{k \in \mathcal{K}}$
with a constraint that $c_k = c_{-k}^*$. We derive the conditions on the coefficients such that the resulting SoS filter follows \eqref{eq:filter_condition}. For this, the Lipschitz constant $\mathcal{L}_g(T_s)$ is computed by using the mean value theorem. For every $t$ and $t+T_s$ within the support of $g(t)$, that is, in the interval $[0, T_g]$ we have that
\begin{align}
\left|  g(t+T_s)-g(t)  \right|  & \leq  \underset{\tau \in [0, T_g]}{\sup} g^{\prime}(\tau) \, T_s \\
& \leq 2\omega_0 T_s \sum_{k=1}^K k|c_k|   = \mathcal{L}_g(T_s).
\label{eq:gt}
\end{align}
While deriving the previous inequality, we used the conjugate symmetry constraint $c_k = c_{-k}^*$.

Substituting \eqref{eq:gt} in \eqref{eq:filter_condition}, we note that \eqref{eq:diff_samples} hold if the filter coefficients and the sampling interval satisfy the following condition:
\begin{align}
	2La_{\max} \alpha_h \omega_0 T_s \left(\sum_{k=1}^K k |c_k|\right) \leq \lambda.
	\label{eq:diff_cond}
\end{align} 
The optimal sampling interval that corresponds to the minimum possible sampling rate in the absence of modulo operation is given as 
\begin{align}
T_{\text{opt}} = \frac{T_d}{(2K+1)}.
\label{eq:Topt}
\end{align}
 If one wants to operate at the minimum possible sampling rate, that is, $T_s = T_{\text{opt}}$ then the filter coefficients must satisfy the inequality 
\begin{align}
\sum_{k=1}^K k|c_k| \leq \frac{\lambda}{2 \omega_0 T_s L a_{\max} \alpha_h} = \frac{(2K+1)\lambda}{4 \pi  L a_{\max} \alpha_h}. 
\label{eq:sos_coeff}
\end{align} 
Alternatively, if the filter coefficients are designed to satisfy a particular criterion, then one may have to sample above the minimum sampling rate. For example, in the presence of noise, the coefficients could be optimally designed to minimize reconstruction accuracy as in \cite{eldar_sos}. 

Next, for the sake of completion of the discussion on the filter design, we consider a simple choice of filter coefficients. Consider an SoS filter with 
\begin{align}
c_k =1, \quad k = -K, \cdots, K.
\label{eq:ck=1}
\end{align}
The dynamic range of $y(t)$ is a function of the coefficients. The maximum value of the filtered output, $\|y(t)\|_{\infty}$, in terms of the filter coefficients and the FRI parameters is derived in the following. 

Using Young's inequality\footnote{Suppose $f \in \mathrm{L}^p(\mathbb{R})$ and $g \in \mathrm{L}^q(\mathbb{R})$, and $\displaystyle p^{-1}+q^{-1} = r^{-1}+1$ with $1 \leq p, q \leq r \leq \infty$, then $\|f*g\|_r \leq \|f\|_p \|g\|_q$.} 
\begin{align}
\underset{t}{\max}|(h*g)(t)| &=	\|(h*g)(t)\|_{\infty}  \leq \|h(t)\|_1 \, \|g(t)\|_\infty \nonumber \\
& \leq \alpha_h  \left(|c_0| + 2\sum_{k=1}^K |c_k|\right).
\label{eq:young}
\end{align}
Next, consider the following relations 
\begin{align}
|y(t)| &= \left|  \sum_{\ell=1}^L a_\ell \, (h*g)(t-t_\ell)\right| \nonumber \\
&\leq \left|(h*g)(t) \right|   \sum_{\ell=1}^L |a_\ell| \leq L a_{\max}\left|(h*g)(t) \right| \nonumber \\
& \leq L a_{\max}\alpha_h  \left(|c_0| + 2\sum_{k=1}^K |c_k|\right) = \|y(t)\|_{\infty},
\label{eq:max_y} 
\end{align}
where the last inequality is derived by using \eqref{eq:young}. Hence for $c_k = 1$, we have that 
\begin{align}
\|y(t)\|_{\infty} = L a_{\max}\alpha_h  (2K+1).
\label{eq:y_max_ck=1}
\end{align}
Let the sampling interval with modulo be
\begin{align}
T_s = \frac{T_{\text{opt}}}{\text{OF}},
\label{eq:Ts_mod}
\end{align} 
where OF is the oversampling factor. Substituting \eqref{eq:ck=1}, \eqref{eq:Ts_mod}, and \eqref{eq:y_max_ck=1} in \eqref{eq:diff_cond} we have
\begin{align}
\text{OF} & \geq 2\pi \frac{L a_{\max} \alpha_h}{\lambda} \frac{K(K+1)}{(2K+1)} \approx \frac{\pi}{2} \frac{\|y(t)\|_{\infty}}{\lambda} , 
\label{eq:of}
\end{align}
for $\|y(t)\|_{\infty} > \lambda$. For $\|y(t)\|_{\infty} < \lambda $, we need not over sample, that is, $\text{OF} = 1$. 

\begin{figure}[!t]
	\begin{center}
		\begin{tabular}{cc}
			\subfigure[]{\includegraphics[width=1.6in]{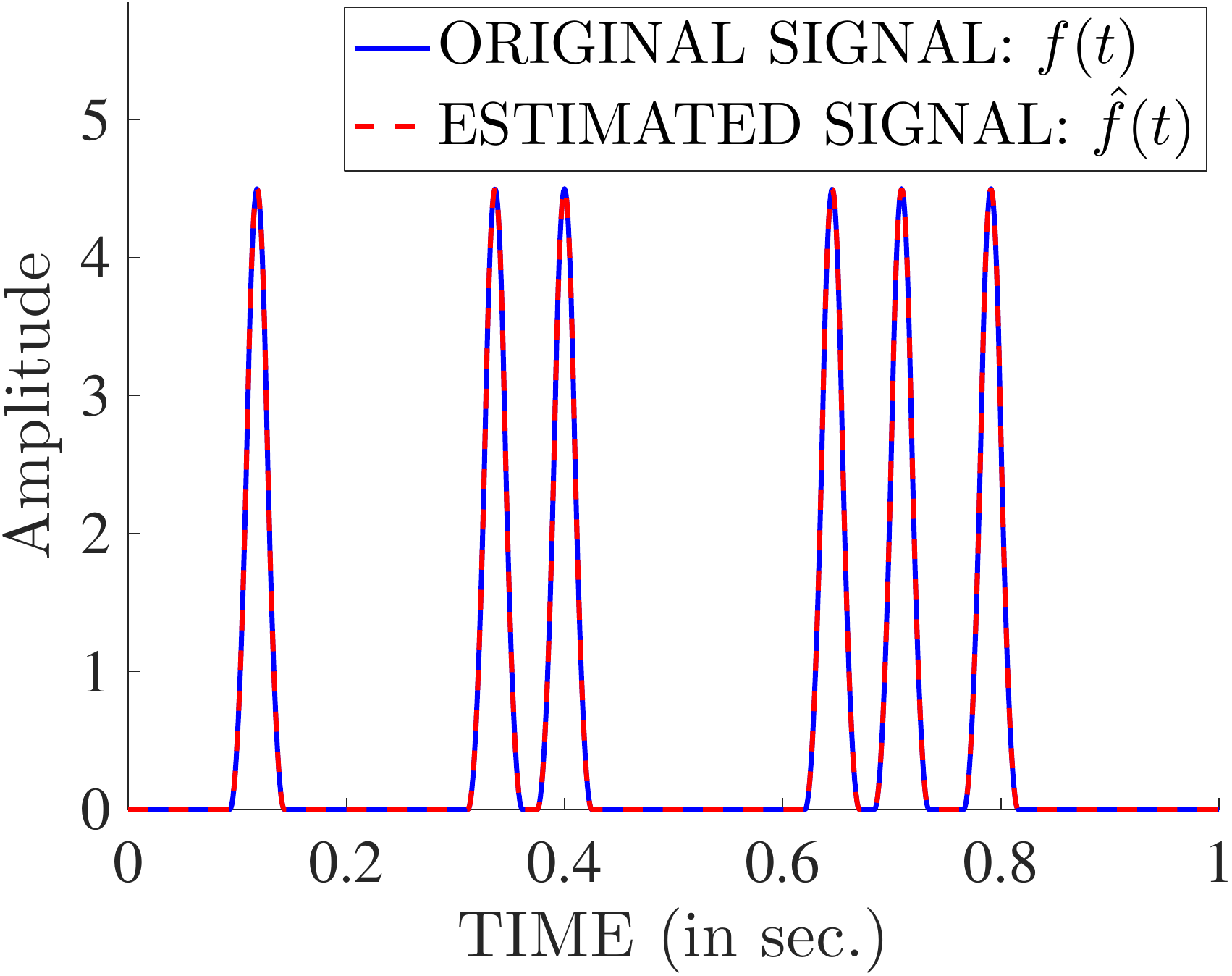}\label{fig:FRIrecoveryA}} 
			\subfigure[]{\includegraphics[width=1.6in]{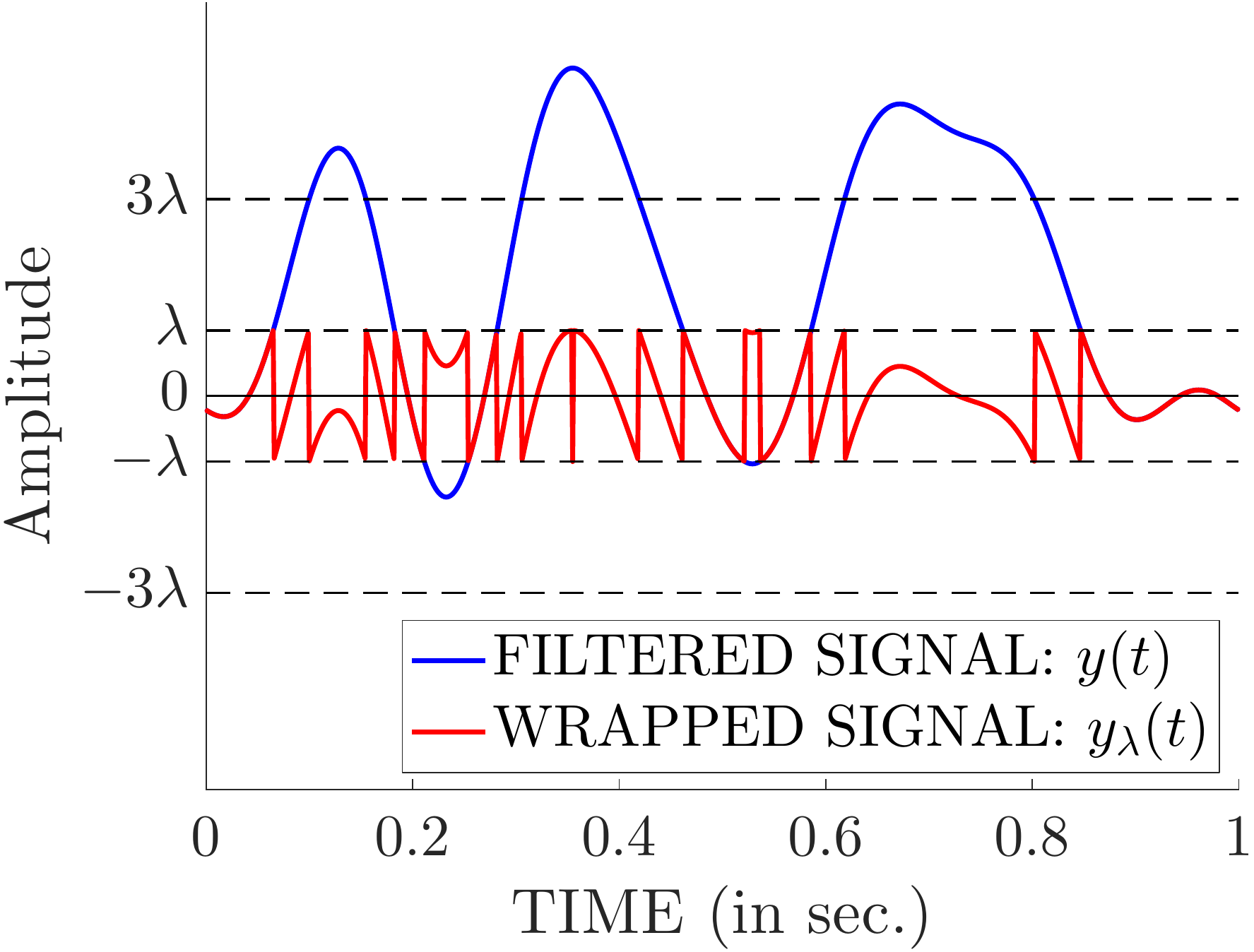}\label{fig:FRIrecoveryB}} 
		\end{tabular}
		\caption{Reconstruction of FRI signal from modulo samples: (a) FRI signal $f(t)$ and its reconstruction $\hat{f}(t)$ where $f(t)$ consists of $L=6$ pulses with $a_{\max} = 6$ and $T_{d}  = 1$; (b) Filtered signal $y(t)$ and its wrapped counterpart $y_{\lambda}(t)$ where $c_k  = 1$ and $\lambda = 0.2 \|y(t)\|_{\infty}$. The signal is reconstructed by using ASAF  where OF = 8 and $2L+1$ samples are used for recovery.}
		\label{fig:FRIrecovery}
	\end{center}
\end{figure}

To summarize, if all the SoS kernel coefficients are set to be unity then by oversampling by a factor computed using \eqref{eq:of} one can ensure that \eqref{eq:y_lip} is satisfied and subsequently the FRI parameters can be estimated using ASAF. To illustrate the accuracy of the algorithm we consider recovery of an FRI signal consisting of $L=6$ pulses with $a_{\max} = 6$ and $T_d = 1$. The pulses $h(t)$ are third-order $\beta$-splines. The signal is filtered by using SoS filter with $c_k = 1$. The filtered signal is wrapped by using $\lambda = 0.2 \|y(t)\|_{\infty}$ and then sampled with an OF = 8 for $K=L$. The FRI parameters are recovered by using ASAF algorithm with $2L+1$ samples of $y_\lambda(t)$. In Fig.~\ref{fig:FRIrecovery}(a), we show that the FRI signal is perfectly recovered. The filtered signal and its modulo version are shown in Fig.~\ref{fig:FRIrecovery}(b) for visualization.

\begin{table*}[!t]
	\footnotesize
	\centering 
	\caption{A Comparison of the Proposed (With positive Amplitudes) Sampling and Reconstruction Method with that in \cite{uls_sparse}, \cite{uls_tsp}, \cite{bhandari2021unlimited}, \cite{uls_romonov}}
	\begin{tabular}{ | m{1.5cm}  | m{2.1cm}  | m{3.2cm}| m{2.6cm} | m{2.6cm}| m{3cm} |}
		\hline &&&&&\\
		~ & L-HoD \cite{uls_sparse} & ~~~~~PBL \cite{bhandari2021unlimited}& ~~~~~~HoD \cite{uls_tsp} & ~~~~~~~LP \cite{uls_romonov} & Our Approach \\ \hline & \multicolumn{5}{ c |}{ }\\  
		& \multicolumn{5}{ c |}{Sampling Setup} \\ \hline 
		Sampling Kernel  &  Periodic LPF (Impractical)  & Periodic LPF/Compact SoS filter & Ideal LPF (Impractical) & Ideal LPF (Impractical) &Compact SoS, practically feasible\\ \hline &&&&&\\
		Minimum Sampling Rate  &  $\frac{2 \pi e (2L+1)}{T_d}$ & ~~$\frac{(2L+1)}{T_d} + \frac{4L}{T_d}\frac{\|y(t)\|_{\infty}}{\lambda}$& ~~~~~$\frac{2 \pi e (2L+1)}{T_d}$& ~~~~~$> \frac{2L}{T_d}$ & $\frac{\pi}{2} \frac{\|y(t)\|_{\infty}}{\lambda} \frac{(2L+1)}{T_d}$\\  
		
		\hline & \multicolumn{5}{ c |}{ }\\  
		& \multicolumn{5}{ c |}{Unfolding and Reconstruction in the Absence of Noise}\\
		
		\hline &&&&&\\
		Min. Number of Samples  &  $2L+1+7\frac{\|y\|_{\infty}}{\lambda}$ & $ 2L+1+ 4L\frac{\|y(t)\|_{\infty}}{\lambda}$ &  All countably infinite samples &  All countably infinite samples& $2L+1$\\

		\hline&&&&&\\
		Unfolding  approach &  Higher-order differences & Out-of-band energy & Higher-order differences & Linear prediction with Chebyshev coefficients& Itoh's method of first-order difference \\ \hline&&&&&\\
		FRI Reconstruction &  Annhilating filter, ignored the constant factor & Not applicable & Not applicable & Not applicable &Modified AF including the constant factor \\ \hline
	\end{tabular}
	\label{table1}	
\end{table*}

\section{A Comparison of Algorithms }
\label{sec:comparision}

In this section, we compare the proposed algorithm with those in \cite{uls_tsp, uls_romonov, uls_sparse, bhandari2021unlimited}. The algorithms in \cite{uls_tsp, uls_romonov} consider an ideal LPF as a sampling kernel. While the unwrapping in \cite{uls_tsp} requires almost 17-times oversampling, unwrapping in \cite{uls_romonov} needs to oversample above the RoI. However, simulation results were not presented for LP approach in \cite{uls_romonov} and, as we show later, in practice, the convergence of the algorithm requires a significantly higher sampling rate. For the rest of the discussion, the algorithms in \cite{uls_tsp} and \cite{uls_romonov} will be refereed as HoD and LP, respectively.

In both the algorithms mentioned above, one is required to use all the countably infinite unwrapped samples to determine the Fourier samples of the FRI signals for subsequent processing. 
In \cite{uls_sparse}, Bhandari et al. considered sampling and \emph{local} reconstruction of FRI signals by using a periodic lowpass filter instead of an ideal LPF. Due to the periodic bandlimited nature of the signal, a finite number of samples measured over a period are sufficient to determine the Fourier samples. Specifically, in \cite{uls_sparse}, the number of time samples and the sampling rate are $2L+1+\frac{\|y\|_{\infty}}{\lambda}$ and $\frac{2 \pi e (2K+1)}{T_d}$, respectively, where $K \geq L$. The oversampling factor is $2\pi e$, which is approximately 17 times higher than the RoI. Here, the number of samples is inversely proportional to $\lambda$, and many measurements are required for low dynamic range ADCs. On the other hand, the sampling rate is independent of the dynamic range of the ADC and always operates at 17-times the RoI. In addition, the authors does not discuss the ambiguity issue that arises while recovering the FRI signals from the Fourier samples. As the method uses local reconstruction together with HoD, we refer to it as local-HoD or (L-HoD).

In \cite{bhandari2021unlimited}, the authors showed that the trigonometric polynomial as in \eqref{eq:ynTs} (or \eqref{eq:y_F}) can be uniquely recovered (up to a constant factor) from its modulo samples provided that $T_s \leq \frac{T_d}{2(K+M+1)}$ where $M$ denotes the number of folding instants incurred in an interval of length $T_d$ while wrapping $y(t)$. All the samples within one time interval,  $2(K+M+1)$, are used for recovering $y(nT_s)$ from its modulo samples. The number of folding instants $M$ is a function of $\frac{\|y(t)\|_{\infty}}{\lambda}$, however, an upper bound on $M$ is not derived in \cite{bhandari2021unlimited}. For a better comparison, we derived an upper bound in the Appendix. We refer to this approach as PBL.

 \begin{figure}[!t]
	\centering
	\includegraphics[width= 2.1in]{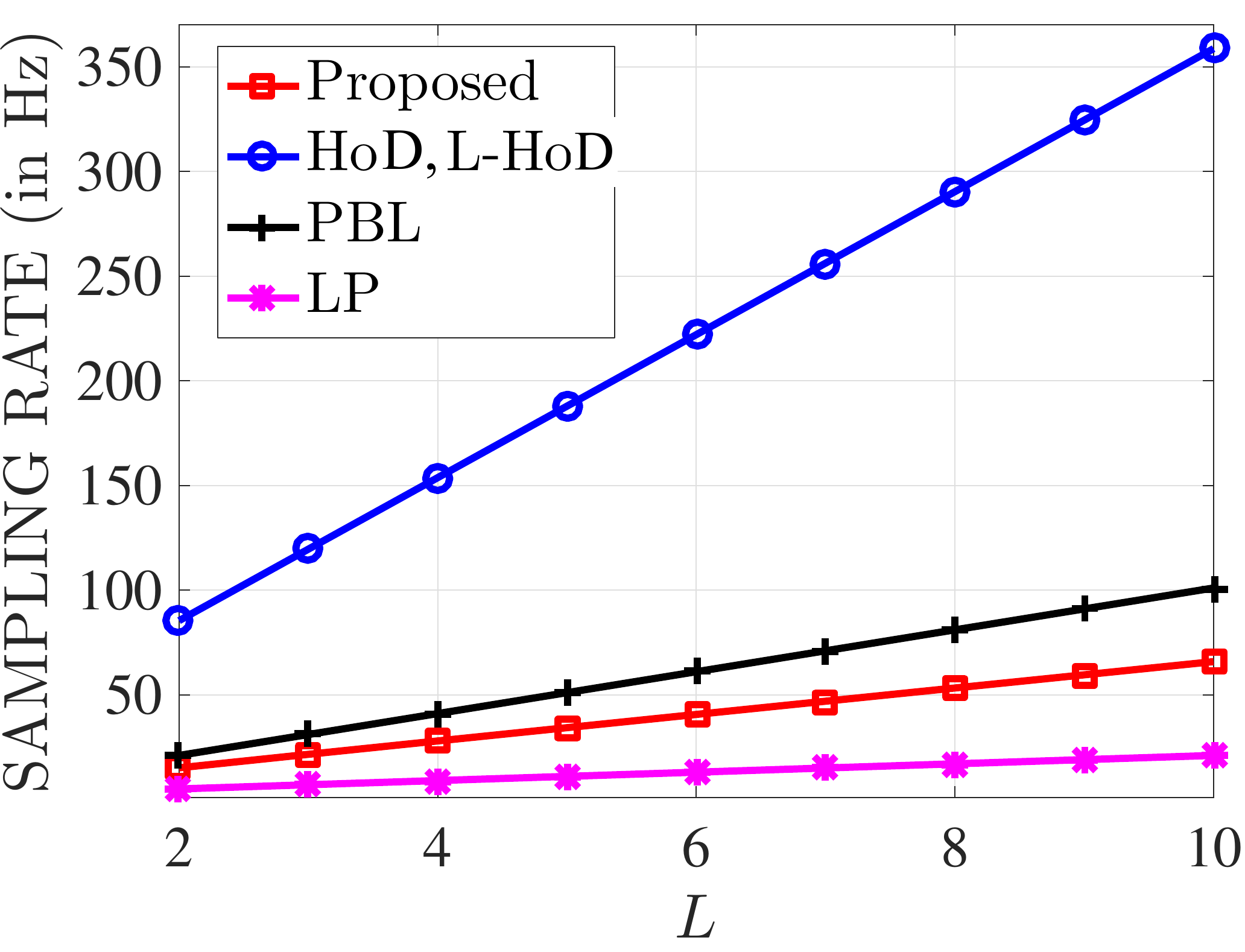}
	\caption{Comparing sampling rates of different algorithms for varying values of $L$ with $\frac{\|y(t)\|_{\infty}}{\lambda} = 2$. The LP approach proposed in \cite{uls_romonov} has lowest sampling rate but requires an infinite number of samples.   }
	\label{fig:comapre_fs}
\end{figure}

 \begin{figure}[!t]
	\centering
		\begin{tabular}{cc}
		\subfigure[]{\includegraphics[width= 1.65in]{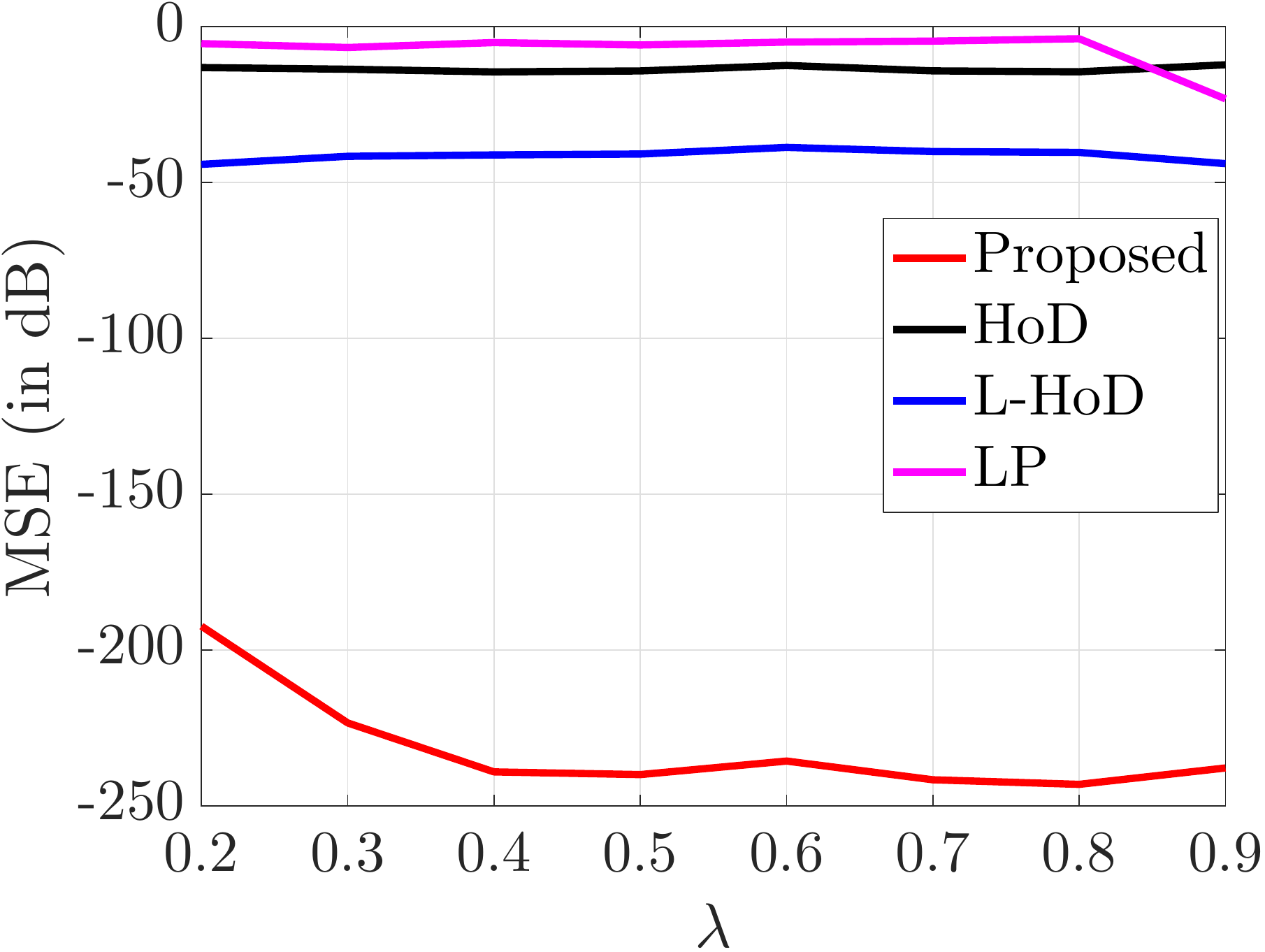}} 
		\subfigure[]{\includegraphics[width= 1.65in]{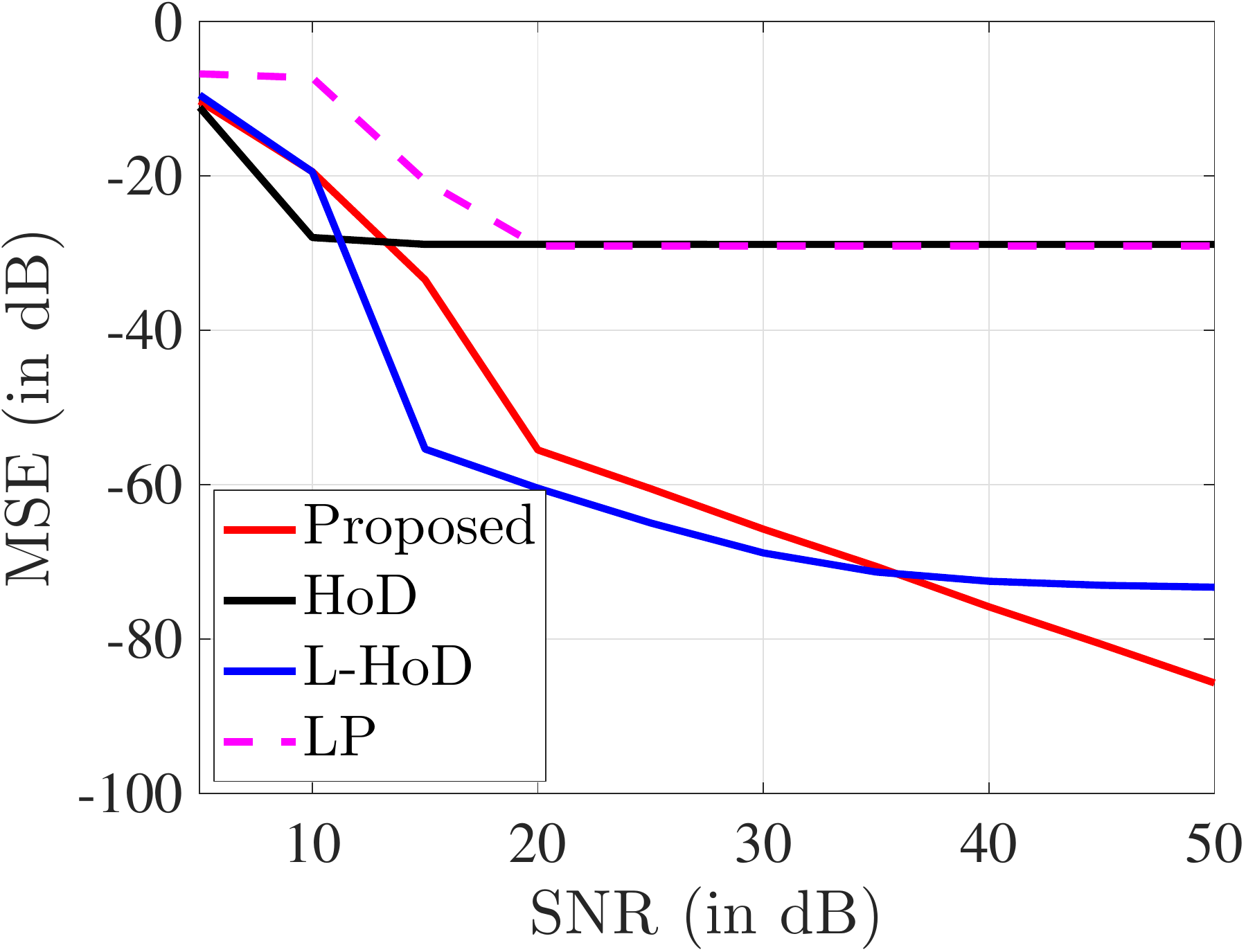}} 
	\end{tabular}
	\caption{Comparison of different methods for $L=3$: (a) MSEs in estimation of time-delays in the absence of noise; (b) MSEs in the estimation of time-delays for different noise levels for $\lambda = 0.3 \|y(t)\|_{\infty}$.  In the absence of noise, the proposed methods result in lowest error. In the presence of noise, the HoD approach (OF = 17) with infinite number of samples has lowest error. The proposed approach operates with OF = 5 and uses only 7 samples.}
	\label{fig:comapre_nonoise}
\end{figure}

A comparison of the algorithms is summarized in Table~\ref{table1}. In these comparisons, we consider the SoS filter with all unit coefficients. We observe that our algorithm operates with the lowest number of measurements, whereas, in the other approaches, either all countably infinite measurements are required \cite{uls_tsp, uls_romonov}, or they are inversely proportional to the dynamic range of the ADC \cite{uls_sparse, bhandari2021unlimited}. Comparing the sampling rates, we observe that for recovering periodic BL signals as in \cite{bhandari2021unlimited} and in our approach, the rate is inversely proportional to $\lambda$, whereas, for BL signals, it is independent of $\lambda$. For better comparison, sampling rates for different algorithms are shown in Fig.~\ref{fig:comapre_fs} where we note that the LP approach \cite{uls_romonov} has the lowest possible rate but requires an infinite number of samples. On the other hand, the proposed approach is second best in terms of sampling rate and requires the lowest number of samples. Note that the approaches in \cite{uls_tsp, uls_romonov,bhandari2021unlimited} are not explicitly designed for FRI signal recovery, which is indicated in the last row of the table. However, while applying these methods for FRI signals, one can use the proposed ASAF approach for recovery.

Next, we compare the algorithms in terms of their accuracy in the estimation of the FRI parameters. Specifically, we consider the mean-squared-error (MSE) in the estimation of time delays which is computed as
\begin{align}
\text{MSE} = \sum_{\ell=1}^L |\tau_\ell -\hat{\tau}_\ell |^2
\end{align}
where $\hat{\tau}_\ell$ is an estimate of $\tau$. First, we compare the algorithms for different values of $\lambda$ without noise. In these simulations, we consider $L=3$, $T_d = 1$, and $h(t)$ to be the Dirac impulse. The time-delays are generated uniformly at random over the interval $(0, 1]$ and are kept constant throughout the experiments. The amplitudes are set to be unity to ensure maximum variation in the filtered signal. The filtered signals are normalized to have a maximum unity amplitude prior to the modulo operation. After unwrapping, we apply ASAF to estimate the time delays. The MSEs in the estimations are shown in Fig~\ref{fig:comapre_nonoise}(a). We observe that the proposed algorithm can estimate the time-delays up to machine precision by using seven samples compared to 239, 120, and 43 samples used by HoD, LP, and L-HoD algorithms. Ideally, both HoD and LP methods require an infinite number of samples. However, in practice, one can use a finite number of samples and the truncation results in a larger error in these algorithms. Both HoD and L-HoD operate at 17 times the rate of innovation while LP is operating with $\text{OF} = 3$. The sampling rate of the proposed algorithm varies with $\lambda$ and in this particular example $\text{OF} = 7.7$ for $\lambda = 0.2$ and  $\text{OF} = 1.7$ for $\lambda = 0.9$.

To evaluate the performance of the algorithms in the presence of noise, we added Gaussian noise with zero mean to the modulo samples prior to unwrapping. In Fig.~\ref{fig:comapre_nonoise}(b), we show the MSEs which are averaged over 1000 independent realizations of noise samples for different signal to noise ratios (SNRs) for $\lambda = 0.2$. Due to the unbounded nature of the noise, the unwrapping algorithms required a higher sampling rate for convergence. In these simulations, we use $\text{OF} = 7$ for the proposed algorithm and $\text{OF} = 17$ for the rest of the three methods. The number of samples used in the proposed approach is 50, whereas 120 samples were used in L-HoD. Both HoD and LP considered 239 samples. As SNR increases, the errors in the proposed method and L-HoD decrease, whereas they stagnate for HoD and LP at higher SNRs due to truncation issues. For SNRs between 10 to 30 dB, the L-HoD method has lower error than the proposed algorithm but at the cost of a higher sampling rate and the number of samples.

For time-delay estimation, we are unable to attain a reasonable MSE while using the algorithm presented in \cite{bhandari2021unlimited} and hence the corresponding results were not included.

\section{Conclusions}
We consider the problem of sampling FRI signals under a modulo framework. We present theoretical guarantees and a practical algorithm for recovering FRI parameters from the modulo samples using an SoS kernel. Theoretical guarantees show that for unique recovery, one needs to sample above the RoI and consider as many samples as the number of unknowns of the FRI signal. The proposed practical algorithms operate at a lower sampling rate and require fewer samples than existing ones. The results enable the design of low-cost and high-dynamic-range ADCs with compactly supported kernels.

\appendix

\section{Upper Bound on $M$ for the Approach in \cite{uls_romonov}}
\label{append:M}
We derive an upper bound on the number of level crossings of a trigonometric polynomial. Consider a $K$-th order trigonometric polynomial 
\begin{align}
	y(t) = \sum_{k = -K}^{K} c_k \, F(k\omega_0) e^{-\mathrm{j}k\omega_0 t}
	\label{eq:y(t)}
\end{align}
which is same as the filtered FRI signal over an interval of length $T_d$. We are interested in finding the maximum number of times $y(t)$ crosses level sets $2\lambda \mathbb{Z}$. In the following, we denote the upper bound as $M$. To this end, let us consider the following lemma.

\begin{lemma}[Level-Crossings of Trigonometric Polynomial]
	\label{lemma:lc_pt}
	Consider a trigonometric polynomial $y(t)$ of order $K$ and time period $T_d$ as in \eqref{eq:y(t)}. Consider an amplitude level $l$ such that $|l| \leq \|y(t)\|_{\infty}$. Then $y(t)$ has a maximum of $2K$ crossings with level $l$ within one time-period $T_d$.
\end{lemma}
\begin{proof}
	The lemma is a direct consequence of the fact that a trigonometric polynomial of order $K$ has maximum $2K$ zeros within one time-period \cite[p. 150]{powell1981approximation}. Determining the level crossings is equivalent to solving for values of $t\in (0, T_d]$ such that $y(t) = l$ or equivalently, $y(t)-l=0$. Since $y(t)-l$ is another trigonometric polynomial of order $K$, it has a maximum of $2K$ zeros within one time-period. 
\end{proof}
The result implies that if $2\lambda n_1 <\|y(t)\|_{\infty}$ for any integer $n_1$ then there will be at max $2K$ foldings corresponding to the level $2\lambda n_1$. Now the maximum number of level-sets $y(t)$ can cross is given by $\displaystyle \left\lfloor \frac{2\|y(t)\|_{\infty}}{\lambda} \right \rfloor$. Hence, $M$ is upper bounded as
\begin{align}
	M \leq \left\lfloor \frac{2\|y(t)\|_{\infty}}{\lambda} \right \rfloor 2K
	\label{eq:Mbound}
\end{align}
The results indicate that the sampling rate as well as the number of samples required for perfect recovery are inversely proportional to $\lambda$. 

Using \eqref{eq:y(t)}, we have that 
\begin{align}
	|y(t)| \leq (2K+1) \, \max_{k \in \mathcal{K}} |c_k| |F(k\omega_0)|.
\end{align}
From \eqref{eq:F}, we note that 
\begin{align}
	|F(\omega)|\leq \|H(\omega)\|_{\infty} L a_{\max} = \|h\|_1 L a_{\max}.
	\label{eq:Fmax}
\end{align}
Hence  
\begin{align}
	\|y(t)\|_{\infty} = (2K+1) c_{\max} \|h\|_1 L a_{\max},
	\label{eq:ymax}
\end{align}
where $c_{\max} = \max_{k \in \mathcal{K}} |c_k|$.
From \eqref{eq:ymax} and \eqref{eq:Mbound}, the sampling rate and the number of samples required for uniquely identifying the FRI signal is on the order of $K^2$ where $K\geq L$.

\bibliographystyle{IEEEtran}
\bibliography{US_biblios,refs,refs2}

\end{document}